\documentclass{article}
\usepackage[a4paper, left=2.5cm, right=2.5cm, top=2.5cm, bottom=3cm]{geometry}
\usepackage[utf8]{inputenc}
\usepackage{url}
\usepackage[usenames,dvipsnames,svgnames]{xcolor}
\usepackage{graphicx}
\graphicspath{{figures/}}

\usepackage{hyperref}

\usepackage{mathtools}
\usepackage{amssymb, amsthm}
\usepackage[ruled]{algorithm2e} 

\usepackage{braket}
\usepackage{thm-restate}

\usepackage{siunitx}
\DeclareSIUnit\volume{vol}
\DeclareSIUnit\time{sec}
\renewcommand{\sec}{\si{\time}}
\newcommand{\vol}{\si{\volume}}

\usepackage{cleveref}
\crefname{lemma}{Lemma}{Lemmas}
\crefname{thm}{Theorem}{Theorems}
\crefname{claim}{Claim}{Claims}

\newcommand{\R}{\mathbb{R}}
\newcommand{\N}{\mathbb{N}}
\newcommand*\diff{\mathop{}\!\mathrm{d}}
\newcommand{\abs}[1]{\left\lvert #1 \right\rvert}
\DeclareMathOperator*{\argmax}{arg\,max}
\DeclareMathOperator*{\argmin}{arg\,min}
\renewcommand{\epsilon}{\varepsilon}

\renewcommand{\l}{\ell} 
\newcommand{\lD}{\ell^{D}} 

\renewcommand{\P}{\mathcal{P}} 
\newcommand{\tauD}{\lceil\tau_e \rceil_\alpha}
\newcommand{\istep}{{\tau^*}} 

\newenvironment{subproof}[1][\proofname]
  {\begin{proof}[#1]}
  {\end{proof}}

\newtheorem{definition}{Definition}
\newtheorem{claim}{Claim}
\newtheorem{theorem}{Theorem}
\newtheorem{lemma}{Lemma}

\usepackage{authblk}
\renewcommand\Affilfont{\itshape\small}
\author[1]{Leon Sering}
\author[1]{Laura Vargas Koch}
\author[2,3]{Theresa Ziemke}
\affil[1 ]{Institute for Operations Research, ETH Z\"urich, R\"amistrasse 101
8092 Z\"urich}
\affil[2 ]{Combinatorial Optimization and Graph Algorithms, TU Berlin, Stra\ss{}e des 17. Juni 136, 10623 Berlin}
\affil[3 ]{Transport Systems Planning and Transport Telematics, TU Berlin, Salzufer 17-19, 10587 Berlin}
{
    \makeatletter
    \renewcommand\AB@affilsepx{: \protect\Affilfont}
    \makeatother
 
    \affil[ ]{Emails}

    \makeatletter
    \renewcommand\AB@affilsepx{, \protect\Affilfont}
    \makeatother

    \affil[ ]{sering@math.ethz.ch}
    \affil[ ]{laura.vargas@math.ethz.ch}
    \affil[ ]{tziemke@vsp.tu-berlin.de}
}

\title{Convergence of a Packet Routing Model to Flows Over Time\footnote{This work was published in Mathematics of Operations Research: \url{https://doi.org/10.1287/moor.2022.1318}.}}

\providecommand{\keywords}[1]
{
  \small	
  \textbf{\textit{Keywords---}} #1
}

\begin{document}
\maketitle

\SetInd{0.75em}{1em} 

\begin{abstract}%
 The mathematical approaches for modeling dynamic traffic can roughly be divided into two categories: discrete
 \emph{packet routing models} and continuous \emph{flow over time models}. Despite very vital research activities on
 models in both categories, the connection between these approaches was poorly understood so far. In this work we build
 this connection by specifying a (competitive) packet routing model, which is discrete in terms of flow and time, and by
 proving its convergence to the intensively studied model of flows over time with deterministic queuing. More precisely,
 we prove that the limit of the convergence process, when decreasing the packet size and time step length in the packet
 routing model, constitutes a flow over time with multiple commodities. In addition, we show that the convergence result
 implies the existence of approximate equilibria in the competitive version of the packet routing model. This is of
 significant interest as exact pure Nash equilibria, similar to almost all other competitive models, cannot be guaranteed
 in the multi-commodity setting.
 
 Moreover, the introduced \emph{packet routing model with deterministic queuing} is very application-oriented as it is
 based on the network loading module of the agent-based transport simulation MATSim. As the present work is the first
 mathematical formalization of this simulation, it provides a theoretical foundation and an environment for provable
 mathematical statements for MATSim.
\end{abstract}

\keywords{flow over time; packet routing; multi-commodity; dynamic equilibrium; convergence of discrete to continuous; discretization error; approximate equilibrium; dynamic traffic assignment}

\section{Introduction.}
Modeling traffic is an essential, but difficult task for which many different approaches have been developed during the
last decades. Some, mainly older approaches model traffic as time-independent \emph{static} flows \cite{PTV2016Visum_16_Manual,rosenthal1973network,wardrop1952road}.
These models are relatively simple and well-understood, e.g.~regarding
existence, uniqueness, efficiency, and structure of equilibria, but they can only model aggregated traffic flows. In
contrast to that, there are time-dependent \emph{dynamic} approaches, which lead to more realistic, but also more
complex models. There exist various mathematical approaches as well as simulation tools in this category. Some consider
discrete time steps \cite{harks2018competitive,horni2016matsim,scarsini2018dynamic,werth2014atomic}, while others
model continuous time \cite{hoefer2011competitive, koch2011nash, scheffler2018nash}. Some model traffic as discrete
travelers (vehicles/packets)  \cite{harks2018competitive, hoefer2011competitive, horni2016matsim, scarsini2018dynamic,
scheffler2018nash, werth2014atomic}, while others model it as continuous splittable flows \cite{adamik2022atomic,frascaria2020algorithms, graf2020dynamic,koch2011nash}.
In this paper we consider two of these approaches: on the one hand, the
intensively studied flow over time model with deterministic queuing \cite{koch2011nash}, which considers continuous time
and continuous flow, and on the other hand, a packet routing model considering discrete travelers and discrete time
steps. The latter is based on the transport simulation MATSim (Multi-agent transport simulation, see
\cite{horni2016matsim}) and it is very similar to the model studied by Werth et al.~\cite{werth2014atomic}.

The flows over time model yields exact user equilibria, called \emph{dynamic equilibria} or \emph{Nash flows over time}
\cite{koch2011nash}. They are guaranteed to exist \cite{cominetti2011existence, cominetti2015existence} and their
structure has been studied intensively
\cite{cominetti2021long,correa2019price,israel2020impact,kaiser2020computation,OlverEtAl2022ContinuityUA,pham2020dynamic,sering2019spillback}.
Even though most of the work has been done for single-origin-destination networks, the existence results and further
structural insights have been transferred to the multi-terminal setting as well
\cite{cominetti2015existence,sering2018multiterminal}; see also~\cite{sering2020diss} for an extended analysis on all
facets of Nash flows over time. Even more general, Meunier and Wagner \cite{meunier2010equilibrium} introduce a broader
class of dynamic congestion games and show the existence of dynamic equilibria.

On the application side, the large-scale, agent-based transport simulation MATSim is used to model real-world traffic
scenarios for transport planning purposes. To achieve this in reasonable time, MATSim simplifies real-world traffic by
discretizing time and aggregating vehicles. A co-evolutionary algorithm is used to compute approximate user equilibria.
Despite the different perspectives of the two approaches, experiments by Ziemke et
al.~\cite{ZiemkeEtAl2020FlowsOverTimeAsLimitOfMATSim} indicate that Nash flows over time are the limits of the
convergence processes when decreasing the vehicle size and time step length in the simulation coherently. This motivates
the question whether the convergence of the flow models can be verified and proven mathematically.

\subsection{Our contribution.} To answer this question, we introduce a multi-commodity packet routing model, which is
basically a mathematical formalization of MATSim's network loading module. This \emph{packet routing model with
deterministic queuing} can be seen as a refinement of the existing packet model studied by Werth et
al.~\cite{werth2014atomic}.

The main contribution of this work consists of the proof that for a fixed commodity-wise flow supply the packet flow in
the discrete model converges to the corresponding flow in the continuous model when the time step length and packet
size go to zero. 

On a high level, the proof works as follows. We assign particles to packets and do an induction over time to show that the travel time difference between particles and corresponding packets is bounded. For this, we consider each arc of the network individually and show how to bound the arrival time deviation between particles and packets at the head of the arc in dependency on the deviation given at the tail.
As traversing an arc takes at least a minimum amount of time, the induction hypothesis can be applied.

Note that this convergence result focuses on the network loading aspect of the models and does not consider any game-theoretical aspects
yet, i.e., the routes for the commodities are fixed. Up to our knowledge, the presented work is the first to prove
convergence of a discrete to a continuous model in the dynamic setting.
 
Additionally, we show that, as a consequence of the convergence proof, the competitive version of the packet routing
model possesses approximate user equilibria. This is particularly surprising since -- up to our knowledge -- the
existence of (approximate) pure Nash equilibria in a multi-commodity setting has not been proven for other similar
competitive packet routing models. Note that \emph{exact} multi-commodity pure Nash equilibria are not guaranteed to
exist in most packet routing models; this is also the case for the model presented here.
 
Our results are of significant interest as they are the first step towards the transfer of established structural
results on dynamic equilibria from flows over time to competitive packet routing games. On the one hand, this provides
the first building block for a mathematical foundation of MATSim, on the other hand, the convergence of the network
loading module of the widely used transport simulation strengthens the relevance of dynamic equilibria in the flows over
time model.

\subsection{Further related work.} The connection between discrete and continuous models in dynamic transportation
settings has been studied before. Otsubo and Rapoport \cite{OtsuboRapoport2008DiscreteVsContDynamicTranspVickrey}
consider the Vickrey model and focus on experimental comparisons and Cantarella and Watling
\cite{CantarellaWatling2016UnifiedDiscreteAndContinuousTimeModel} compare discrete and continuous time models by
presenting mathematical tools and applications and unifying both in a single model.

In static routing games, convergence has been shown very recently by Cominetti et al.~\cite{cominetti2020convergence}.
They prove convergence of equilibria in routing games from two perspectives, on the one hand, when player sizes
decreases, and on the other hand, when the probability with which players participate in the game decreases.

For flows over time a connection between discrete and continuous optimization problems was established by Fleischer and
Tardos~\cite{fleischer1998efficient} in 1998. They transferred optimization algorithms for discrete-time flow models to
a flow model with continuous time.

The competitive packet routing model presented in this paper is similar to existing packet routing models with first in
first out (FIFO) rules; see~\cite{caoatomic,Ismaili2017RoutingGamesOverTimeFIFO,scarsini2018dynamic,werth2014atomic}.
However, each model has a different focus. Werth et al.~\cite{werth2014atomic} consider complexity questions and the
price of anarchy for several different objective functions: arrival times and bottleneck costs as players' objectives
and sum of completion times as well as makespan as social cost. While
Ismailli~\cite{Ismaili2017RoutingGamesOverTimeFIFO} analyzes the computational complexity of computing a best response
and a Nash equilibrium, Scarsini et al.~\cite{scarsini2018dynamic} analyze the price of anarchy of a packet routing
model with periodic inflow rates. Finally, Cao et al.~\cite{caoatomic} analyze the existence of subgame perfect
equilibria.

\section{Packet routing model with deterministic queuing.} \label{sec:model} 
In this section, we formally introduce our packet routing model, which is motivated by the network loading model of
MATSim. Packets with individual origins and destinations travel in discrete time steps through a network, where arcs
have a transit time and a capacity. After traversing an arc, packets might be delayed at bottlenecks, where they have to
wait with other packets in a point queue. In order to enter the next arc, packets from different incoming arcs are
merged as smoothly as possible, similar to the zipper-style merging from real traffic. The length of a time step and the packet
size are given by some discretization parameters $(\alpha, \beta)$, which are crucial to show convergence later on. Note
that our model is similar to the model of Werth et al.~\cite{werth2014atomic}. The only differences are the tie-breaking rules in the merging process and that we allow fractional capacities.  
 
\subsection{Preliminaries.} 
\paragraph{Units.} Since we consider convergence later on, we keep the time and packets size flexible. As help for
orientation, we introduce some units for time and for the size of the packets. We fix two base units, namely $\sec$ for
a base unit of time, and $\vol$ for the base unit of packet volume. Note, however, that the used variables are all
considered to be unitless and we use these base units only for intuition.
 
\paragraph{Network.} A network is given by a directed graph $G = (V, E)$, where every arc is equipped with a
transit time $\tau_e > 0$ and a capacity $\nu_e > 0$. The transit time denotes how long it takes to traverse an arc,
measured in $\sec$s. The capacity restricts the flow rate that is allowed to leave the arc. Hence, it is measured in
flow volume per time unit, i.e., $\vol / \sec$.

\paragraph{Discretization.} For our packet routing model we specify two discretization parameters $\alpha,
\beta>0$, one for the time step length and one for the packet sizes. In other words, one time step has a duration of
$\alpha \, \sec$ and each packet has a volume of $\beta \, \vol$. Since the point in time corresponding to a time step
$t \in \N_0$ is given by $\alpha t$, the set of all these points in time is given by $\Theta_{\alpha} \coloneqq  \alpha
\N_0 = \set{0, \alpha, 2 \alpha, \dots}$. In addition, we use rounding brackets in order to round to the previous or
next element of $\Theta_{\alpha}$:
\[\lfloor x \rfloor_\alpha \coloneqq \max \set{\theta \in \Theta_\alpha | \theta \leq x} \quad \text{ and } 
\quad \lceil x \rceil_\alpha \coloneqq \min \set{\theta \in \Theta_\alpha | \theta \geq x}.\]
Analogously, this notation is also used for $\beta$, i.e.,
\[\lfloor x \rfloor_\beta \coloneqq \max \set{\phi \in \beta \N_0 | \phi \leq x} \quad \text{ and } 
\quad \lceil x \rceil_\beta \coloneqq \min \set{\phi \in \beta \N_0  | \phi \geq x}.\]
 
\paragraph{Discretized network.} Given a network and discretization parameters, we normalize the arc properties as
follows. The number of time steps required for a packet to traverse an arc is denoted by $\hat \tau_e \coloneqq
\frac{\tauD}{\alpha} \in \N$. The maximum number of packets that are allowed to leave an arc at each time step is given
by $\hat \nu_e \coloneqq \frac{\nu_e \cdot \alpha}{\beta}$. Note that $\hat \nu_e$ does not need to be an integer. The
fractional remainder is transferred to the next time step, as described later.

\paragraph{Packets.} Given a network $G$ and discretization parameters $\alpha$ and $\beta$, we consider $n$ packets
of volume~$\beta$ with a path $P_i$ connecting an origin-destination pair $(o_i, d_i) \in V^2$ and a release time $r_i
\in \Theta_\alpha$ for all $i \in N \coloneqq \set{1, 2, \dots, n}$. The release time step is hence given by $\hat r_i
\coloneqq \frac{r_i}{\alpha} \in \N_0$.

\subsection{Network loading.} Given a network, discretization parameters, and a set of packets it is necessary to
determine the movement of the packets through the network. We do this by specifying the arc dynamics and the node
transitions algorithmically, which combined determines the position of every packet over time. In every discrete time
step, we first consider the dynamic of all arcs. This primarily means determining the packets that leave each arc.
Afterwards, the node transition is executed for all nodes, moving the packets to the next arc on their respective path.
This alternating process is repeated for every time step until every packet has reached its destination. Note that this
happens in finite time, as packets travel along simple paths. Since all transit times are strictly positive, a packet
can only enter one arc per time step. This justifies processing all arc dynamics first while computing the node
transitions afterwards at each time step.
 
\begin{figure}
\centering
\includegraphics{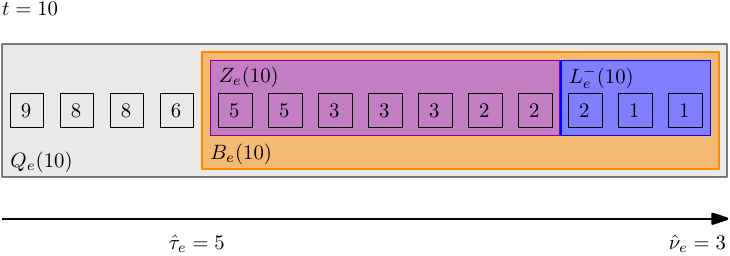}
\caption{Example of an arc queue $Q_e$ at time step $t = 10$. Different parts are highlighted: Packets in the
buffer~$B_e$, leaving packets $L_e^-$ and waiting packets $Z_e$. Within each packet, the arc entrance time step is
denoted.}
\label{fig:arc_packet}
\end{figure}

\paragraph{Arc queue.} We consider every arc $e$ of the network to be a packet queue $Q_e$ denoted as an ordered
finite sequence $Q_e(t)$ of packets for every time step $t \in \N_0$. In addition, packets are labeled with the time
step at which they have entered~$e$. Note that at any time step the ordering of all packets that are currently on this
arc respects the arc entrance times. In other words, the queues operate by the first in first out (FIFO) principle. How
ties are broken for packets that enter an arc at the same time step is specified by the node transition algorithm
given below. In the following, we denote different parts of the arc queue, which are also illustrated in
\Cref{fig:arc_packet}.

\paragraph{Buffer.} At every time step $t$ the \emph{buffer}~$B_e(t)$ is defined as the sequence of packets on
arc~$e$, that have been in $Q_e$ long enough time to possibly leave $e$ at time step~$t$. More formally, we define
$B_e(t)$ to be the maximal sequence of packets that satisfy
\begin{enumerate}
\item $B_e(t)$ is a suffix of $Q_e(t)$,
\item the arc entrance time step of all packets in $B_e(t)$ is at most $t - \hat \tau_e$.
\end{enumerate}

\paragraph{Current capacities.} Not necessarily all packets in the buffer will be able to leave the arc at time
step~$t$, as their volume might exceed the current arc capacity. Since the discretized capacity $\hat \nu_e$ can be
non-integer and surely the number of packets leaving the arc has to be integer, the remainder is passed on to the next
time step. For this reason we introduce the current packet capacity $\hat \nu_e(t)$. It is initialized by $\hat \nu_e(0)
\coloneqq \hat \nu_e$ and then defined iteratively by
\begin{equation}
\label{eq:capacity}
\hat \nu_e(t) \coloneqq \begin{cases}
\hat \nu_e  & \text{ if } \abs{B_e(t - 1)} \leq \hat \nu_e(t - 1),\\
\hat \nu_e + \hat \nu_e(t - 1) - \lfloor \hat \nu_e(t - 1)  \rfloor & \text{ else.}
\end{cases}
\end{equation}

Note that $\lfloor \hat \nu_e(t - 1)  \rfloor$ describes the number of packets that will be allowed to leave arc $e$ at
time step $t-1$ as we specify in the next paragraph. Hence, $\hat \nu_e(t - 1) - \lfloor \hat \nu_e(t - 1)  \rfloor$ is
the unused remaining capacity of the last time step, in the case that the number of buffering packets in $B_e(t - 1)$
exceeds the previous capacity $\hat\nu_e(t - 1)$.

\paragraph{Leaving packets.} At every time step, the algorithm determines at every arc $e$ a list of leaving
packets~$L_e^-(t)$, which is an ordered list of packets in the buffer that will leave the arc at this time step.
Formally, $L_e^-(t)$ is the maximal sequence of packets that satisfies the following two conditions:
\begin{enumerate}
\item $L_e^-(t)$ is a suffix of $B_e(t)$,
\item the number of packets in $L_e^-(t)$ does not exceed the current capacity $\hat \nu_e(t)$.
\end{enumerate}
In other words, these are the first $\lfloor \hat \nu_e(t) \rfloor$ packets in line which have spent enough time on $e$
to traverse it. Clearly, this $L_e^-(t)$ is unique and it can be determined algorithmically by considering one packet
after the other in $B_e(t)$.

\paragraph{Waiting packets.} The waiting packets $Z_e(t)$ at time step $t$ are exactly the packets in the buffer
$B_e(t)$ that are not leaving at time step $t$, i.e., that are not contained in $L_e^-(t)$. Following the naming of the
flow over time model we refer to $Z_e(t)$ as \emph{(waiting) queue}.

\paragraph{Released packets.} We have an additional list $L_v^-(t)$ for each node $v$, which contains the packets
that are released into the network through $v$ at time step $t$. In other words,
\[L_v^-(t) \coloneqq \set{ i \in N | o_i = v \text{ and } \hat r_i = t}.\] 
Here, the ordering is given by the packet index. 

\begin{figure}
\centering
\begin{minipage}{.48\textwidth}
\centering
\includegraphics[width=.68\linewidth]{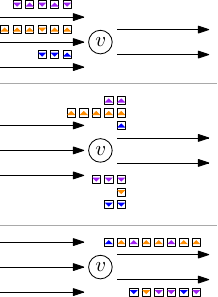}
\caption{The node transition in three steps: leaving packets of each incoming arc are depicted on the left. The small
arrows indicate onto which outgoing arc the packets will be moved. As a first step, the packets get grouped by desired
outgoing arcs (middle). Finally, the packets are merged according to the priority procedure (right side). Here, ties are
broken from the top incoming arc to the bottom incoming arc.}
\label{fig:node_transition}
\end{minipage} \hfill
\begin{minipage}{.48\textwidth}
\centering
\includegraphics[width=\linewidth]{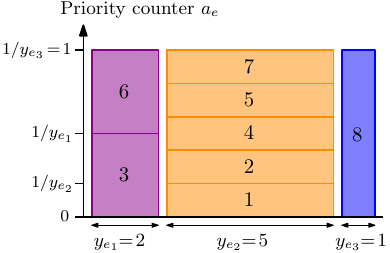}
\caption{Visualization of the merging process with priority counter. We consider the merging process of the upper
outgoing arc of \Cref{fig:node_transition}. For each incoming arc, we draw the packets as a stack of rectangles each of
area $1$ with a width of $y_{e_i}$. The height is then describing the increment of the priority counter. Whenever a
packet is selected by the algorithm, the height of the upper arc of the rectangle corresponding to the packet equals the
priority counter. As packets with a low priority counter are selected first, we can order the packets from the bottom to
the top (always considering the upper edge of the rectangle). Note that ties are broken in favor of the arc with the
lower index.}
\label{fig:priority_counter}
\end{minipage}
\end{figure}

\paragraph{Node transition.} At each time step every leaving packet is moved from its current arc to its subsequent
arc according to its path. During this process, the ordering of the packets is preserved. Whenever packets from multiple
arcs enter into the same arc they are merged in a zipper-style way similar to real traffic. This means that packets are
merged separately on each arc proportional to the number of packets sharing the same previous arc; see
\Cref{fig:node_transition} for an example. The precise procedure can be found in \Cref{alg:node_transition} and is
explained in the following.

\begin{algorithm}
\caption{Node transition at node $v$ at a fixed time step $t$ (the parameter $t$ is omitted here).}
\label{alg:node_transition}
\KwIn{leaving packets $L_{e'}^-$ of all incoming arcs $e' \in \delta_v^-$; \\
\phantom{\textbf{Input: }}list of packets $L^-_v$ that want to enter the network at $v$ at the current time step.}
\KwOut{sorted list $L_e^+$ of packets entering arc $e$ for all outgoing arcs $e \in \delta_v^+$.}
$L_e^+ \leftarrow ()$ for all $e \in \delta_v^+$\\
\For{each $e \in \delta_v^+$}{
  \vspace{0.5em}\tcp{initialization of incoming arcs:}
  \For{each $e' \in \delta_v^-$}{
    $Y_{e'} \leftarrow$ ordered sublist of packets $i$ from $L_{e'}^-$ with $(e', e)$ on path $P_i$\\
    \If{$\abs{Y_{e'}} \geq 1$}{
      $y_{e'} \leftarrow \abs{Y_{e'}}$\quad \tcp{number of packets going from $e'$ to $e$} 
      $a_{e'} \leftarrow \frac{1}{y_{e'}}$ \quad \tcp{priority counter (small means higher priority)}
    }
  }
  $A \leftarrow \set{e' \in \delta_v^- | \abs{Y_{e'}} \geq 1}$ \quad \tcp{arcs that contain packets moving into $e$}
  \vspace{0.5em}\tcp{initialization of released packets:}  
  $Y_v \leftarrow$ ordered sublist of packets $i$ from $L^-_v$ where $e$ is the first arc of path $P_i$\\
  \If{$\abs{Y_v} \geq 1$}{
    $y_v \leftarrow \abs{Y_v}$ \quad \tcp{number of packets released into $e$}
    $a_v \leftarrow \frac{1}{y_v}$ \quad \tcp{priority counter}
    $A \leftarrow A \cup \set{v}$ \quad \tcp{$v$ is treated as an incoming arc}
  } \vspace{0.5em}
  \tcp{transition:}
  \While{$A \neq \emptyset$}{
    $e^* \leftarrow \argmin_{e' \in A} a_{e'}$ \quad \tcp{ties are broken arbitrarily (but deterministically)}
    $L_e^+.append(Y_{e^*}.pop())$ \quad \tcp{remove first packet from $Y_{e^*}$, add it to end of $L_e^+$}
    $a_{e^*} \leftarrow a_{e^*} + \frac{1}{y_{e^*}}$\\
    \If{$\abs{Y_{e^*}} = 0$}{
      $A \leftarrow A \setminus \set{e^*}$
    }
  }
}
\Return{$(L_e^+)_{e \in \delta_v^+}$}

\end{algorithm}
The algorithm considers a node $v$ at a fixed time step $t$ and describes how packets traverse this node. The basic idea
is to consider each outgoing arc $e$ individually. All incoming arcs $e'$ that contain at least one leaving packet that
wants to continue its journey on arc $e$ are collected in the set $A$. Here, packets that enter the network at node $v$
at the current time step are treated as if they would be leaving packets of an additional arc.
The packets should merge as smoothly as possible. Hence, in order to obtain a good ordering of packets entering $e$, we
define a priority counter for all incoming arcs $e'$ that operates proportionally to the number of packets $y_{e'}$
transiting from $e'$ to $e$ at the current time step. Initially, these counters are set to $\frac{1}{y_{e'}}$ and the
incoming arc with the lowest counter (highest priority) can send the next packet. Afterwards, the lowest counter is
increased by $\frac{1}{y_{e'}}$ and if $e'$ does not contain any further leaving packets it is removed from $A$. A
visualization of the priority counter for the merging process of the top outgoing arc in the example of
\Cref{fig:node_transition} is given in \Cref{fig:priority_counter}.
 Note that in the end all leaving packets are transferred onto their next arc and that this merging procedure is mainly
 important for the ordering on the outgoing arcs.

\begin{algorithm}
\SetKwFunction{NodeTransition}{NodeTransition}
\caption{Network loading.}
\label{alg:network_loading}
\KwIn{discretized network $(V, E, (\hat \nu_e)_{e \in E}, (\hat \tau_e)_{e \in E})$; set of packets $N$ with $(o_i, d_i, P_i, \hat r_i)_{i \in N}$.}
\KwOut{arrival time steps $(\hat c_i)_{i \in N}$.}
$t \leftarrow 0$\\
$\hat c_i \leftarrow \infty$ for all $i \in N$\\
$Q_e(0) \leftarrow ()$ for all $e \in E$\\
$\hat \nu_e(0) \leftarrow \hat \nu_e$ for all $e \in E$\\[.5em]
\While{$\exists \; i \in N$ with $\hat c_i = \infty$}{
  determine $L_e^-(t)$ as defined in paragraph \emph{leaving packets} for all $e \in E$\\
  \For{each $i \in N$ with $i \in L_e^-(t)$ and $e \in \delta_{d_i}^-$}{
    $\hat c_i \leftarrow t $ \quad \tcp{packet $i$ has reached its destination}
  }
  \For{each $v \in V$}{
    $L_v^-(t) \leftarrow $ list of all packets $i \in N$ with $\hat r_i = t$ and $o_i = v$ ordered by packet index\\
    $(L_e^+(t))_{e \in \delta_v^+} \leftarrow \NodeTransition((L_{e}^-(t))_{e \in \delta_v^-}, L_v^-(t))$ \\
  }
  \For{each $e \in E$}{
    set arc-entrance time step to $t$ for each packet in $L_e^+(t)$\\
    $Q_e(t + 1) \leftarrow Q_e(t) - L_e^-(t) + L_e^+(t)$ \tcp{remove leaving \& append entering packets}
    update the current capacity $\hat \nu_e(t + 1)$ according to \Cref{eq:capacity}  
  }
  $t \leftarrow t + 1$\\
}
\Return{$(\hat c_i)_{i \in N}$}
\end{algorithm} 

\paragraph{Network loading.} The loading of the network can be defined algorithmically as follows. For every time
step, we first consider all arcs and determine the leaving packets. Next, all leaving packets are moved to the next arc
of their path according to the node transition. Finally, the queues and the current capacities are updated. This is
described formally in \Cref{alg:network_loading}.
 
\subsection{Competitive packet routing game.} In this section, we consider the model from a game-theoretical perspective,
which means that every packet is considered to be a player who aims at arriving as early as possible at her destination.

\paragraph{Strategies and costs.} We are given a discretized network $(V, E, (\hat \nu_e)_{e \in E}, (\hat
\tau_e)_{e \in E})$ and a set of packets~$N$, each packet $i \in N$ equipped with an $(o_i,d_i)$-pair and a release time
$\hat r_i$.  Note that we assume that $d_i$ is reachable from $o_i$ for all $i \in N$. By considering the packets as
players and denoting the set of all simple $o_i$-$d_i$-paths as the strategy set of player $i$, we obtain a competitive
packet routing game.

For every path profile $\pi = (P_i)_{i \in N}$ the network loading given by \Cref{alg:network_loading} determines the
arrival time steps $(\hat c_i(\pi))_{i \in N}$. The arrival time given by $c_i(\pi) \coloneqq \alpha \cdot \hat
c_i(\pi)$ is the cost of player $i$, which she aims to minimize.

\paragraph{Equilibria.} We consider exact pure Nash equilibria in this model. These are states in which no player can
arrive strictly earlier by unilaterally changing her path. This equilibrium concept has been considered in related
models before \cite{caoatomic,werth2014atomic}. 

In our model -- similar to other packet routing models with FIFO rule -- pure Nash equilibria do not exist in general. This follows from the multi-commodity example presented by
Ismaili \cite{Ismaili2017RoutingGamesOverTimeFIFO}, which we adapt to our model here.

\pagebreak[4]

\begin{restatable}{proposition}{thmnoequilibria} \label{thm:no_equilibria}
There are competitive packet routing games in our model for which no pure Nash equilibrium exists. 
\end{restatable}

\begin{proof} \label{proof:no_equilibria}
Consider a game with six players played on the discretized network depicted in \Cref{fig:NoPNE} with $\alpha = \beta =
1$. We have two main players, a pursuer $1$ going from $o_P$ to $d_P$ and an evader $2$ going from $o_E$ to $d_E$. Each
of these two players has exactly two paths to choose from, which we denote by top and bottom. Moreover, we have four
additional players $3, \dots, 6$ with origins $o_3=o_4=o_E$ and $o_5=o_6=o_P$ and destinations $d_3, \dots d_6$. Note
that these players have a unique path, and hence only a single strategy. Their purpose is to transfer the information
between the main players.

\begin{figure}[b]
\centering
\includegraphics{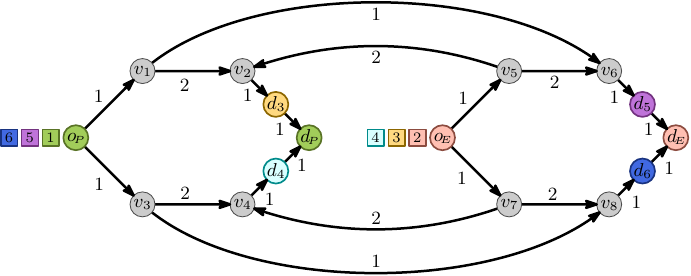}
\caption{Network of a game without pure Nash equilibrium. There are six players, indicated by different colors. The
transit times are depicted beside the arcs and all arcs have a capacity of $1$.}
\label{fig:NoPNE}
\end{figure}

All players have a release time of $0$ and are ordered by their index in the beginning. In a tie-break in the merging
procedure the four long arcs ($v_1v_6, v_5v_2, v_7v_4, v_3v_8$) are prioritized. Note that all arc capacities are $1$,
which implies that also the current capacities of all arcs are constant $1$ (as there are no remaining capacities).
Therefore, at most one packet can leave an arc per time step. Furthermore, if two packets enter into an arc at the
same time the priority counter will always end up in a tie (both counters are $1$), which means that the packet coming
from the prioritized arc (long arc) will always be transferred first.

First, we consider the network loading of the state where both the pursuer and the evader choose their respective top
path. In this case, the evader delays player~$3$ at the end of arc $o_E v_5$ by one time step and the pursuer delays
player~$5$ at $o_P v_1$. Thus, the pursuer enters arc $v_2d_3$ before player $3$ (as she arrives there at time $3$ and
player $3$ arrives at time $4$). Hence, the pursuer reaches $d_P$ at time $5$. On the other side, the evader is delayed
by player $5$ when leaving $v_6d_5$ (since both players have entered the arc at the same time, but due to the tie-break,
player $5$ is in front of the evader in the arc queue). Thus, the evader reaches $d_E$ at time $6$.

Now consider the state where the pursuer chooses the top path and the evader chooses the bottom path. In this case, the
evader delays player $4$ at the end of arc $o_E v_7$ by one time step and the pursuer still delays player $5$ at $o_P
v_1$. Thus, the pursuer enters arc $v_2d_3$ at the same time as player $3$. By the tie-breaking rule of the arcs, player
$3$ is inserted in the arc queue of $v_2d_3$ before the pursuer, and thus, delays her when leaving the arc. This leads
to an arrival time of the pursuer of $6$. The evader reaches node $v_8$ at time $3$ when player $6$ passed already and
can thus arrive without delay at time $5$ in $d_E$.

Since the game is highly symmetric this analysis gives us that the pursuer prefers to play strategy profiles top/top and
bottom/bottom, while the evader prefers strategy profiles top/bottom and bottom/top. Since this is a classical matching
pennies situation the game has no pure Nash equilibrium.
\end{proof}

\paragraph{$\epsilon$-Equilibria.} Since the existence of pure Nash equilibria cannot be guaranteed, we consider
approximate pure Nash equilibria. In these states players choose nearly optimal strategies, i.e., they cannot improve by
more than $\epsilon$ by changing their paths.

\medskip
\begin{definition}
For $\epsilon > 0$ an $\epsilon$-equilibrium is a strategy profile $\pi = (P_i)_{i \in N}$ such that for every player 
$i \in N$ and every simple $o_i$-$d_i$-path $P'_i$ it holds that
\[ c_i(\pi) \leq  c_i(P_{-i},P_i') + \epsilon.\]
Here, $(P_{-i},P_i')$ denotes the strategy profile where all players except player $i$ act as in $P$ and only player $i$ plays a different strategy, namely $P_i'$. 
\end{definition}
\medskip

In the example of $\Cref{thm:no_equilibria}$ every state is a $1$-equilibrium but there do not exist any
$\epsilon$-equilibria for $\epsilon < 1$. Thus, for a given game and a given $\epsilon$ it is not obvious whether there
exists an $\epsilon$-equilibrium or not. In \Cref{sec:game_theory}, however, we prove that for any network (with given
supply rates, which are defined later on) and any given $\epsilon > 0$ we can find a discretization such that an
$\epsilon$-equilibrium exists.

\section{Flows over time.} \label{sec:flowsovertime} 
In order to be able to analyze the connection between the packet routing model and flows over time in the following
sections, we briefly introduce the multi-commodity flow over time model which is studied in
\cite{cominetti2015existence,graf2020dynamic,sering2020diss,sering2018multiterminal}. This model extends the network
flow concept by a continuous-time component. Each infinitesimally small flow particle needs a fixed time to traverse an
arc before it reaches a bottleneck given by the capacity rate of the arc. Whenever the flow rate exceeds this capacity a
queue builds up, which operates at capacity rate.

\subsection{Model.} A flow over time \emph{instance} consists of a network $(V, E, (\nu_e)_{e \in E}, (\tau_e)_{e \in
E})$ together with a finite set of commodities~$J$, where each commodity~$j \in J$ is equipped with a set of particles
$M_j = [0, m_j]$, $m_j \in \R_{> 0}$, an origin-destination-pair $(o_j, d_j)$ and an integrable and bounded \emph{supply
rate} function $u_j \colon [0, \infty) \to [0, \infty)$. The inflow rate function $u_j$ must have bounded support and
must satisfy $\int_0^{\infty} u_j(\xi) \diff \xi = m_j$.

A multi-commodity flow over time is represented by a set of functions $f = (f^+_{j,e}, f^-_{j,e})_{j \in J, e \in E}$
that are bounded and locally integrable. These functions describe the rate with which the flow of commodity~$j$ enters
and leaves arc $e$ at every point in time.  Since flow does not enter the network before time $0$, we set
$f^+_{j,e}(\theta)= f^-_{j,e}(\theta)=0$ for $\theta <0$ and all $j \in J$, $e\in E$. We denote the \emph{cumulative flow} by
\[F^+_{j,e}(\theta) \coloneqq \int_{0}^{\theta} f_{j,e}^+(\xi) \diff \xi \quad \text{ and } 
\quad F^-_{j, e}(\theta) \coloneqq \int_{0}^{\theta} f_{j, e}^-(\xi) \diff \xi.\]
Furthermore, we denote the \emph{total (cumulative) flow}, which is the sum over all commodities, by
\[f_e^+(\theta) \coloneqq \sum_{j \in J} f_{j,e}^+(\theta), \quad f_e^-(\theta) 
\coloneqq \sum_{j \in J} f_{j, e}^-(\theta), \quad F_e^+(\theta) \coloneqq \sum_{j \in J} F_{j,e}^+(\theta) 
\quad \text{ and } \quad F_e^-(\theta) \coloneqq \sum_{j \in J} F_{j, e}^-(\theta).\]

In order to describe the flow dynamics in the deterministic queuing model, several conditions need to be fulfilled for
almost all $\theta \in [0, \infty)$. First of all, we need commodity-wise flow conservation at every node at every point
in time:
\begin{equation} \label{eq:flow_conservation}
\sum_{e\in \delta^+_v} f_{j, e}^+(\theta) - \sum_{e \in \delta^-_v} f_{j, e}^-(\theta) \begin{cases}
= 0 & \text{ for all } v \in V\setminus \set{\!o_j, d_j\!}\!, \\
= u_j (\theta) & \text{ for } v = o_j, \\
\leq 0 & \text{ for } v = d_j.
\end{cases} 
\end{equation}

Let the total flow volume in the queue at time $\theta$ be denoted by $z_e(\theta)$. For particles that enter arc $e$ at time
$\theta$ we obtain a \emph{waiting time} $q_e(\theta)$ and an \emph{exit time} $T_e(\theta)$. Formally, they are given
by
\[z_e(\theta) \coloneqq F_e^+(\theta - \tau_e) - F_e^-(\theta), \qquad 
q_e(\theta) \coloneqq \frac{z_e(\theta + \tau_e)}{\nu_e} \quad \text{ and } \quad 
T_e(\theta) \coloneqq \theta + \tau_e + q_e(\theta).\]

As the queue operates at capacity rate whenever we have a positive queue, we obtain the following condition on the total
outflow rate:
\begin{equation}\label{eq:totaloutflow}
f_e^-(\theta) = \begin{cases}
\nu_e & \text{ if } z_e(\theta) > 0, \\
\min\Set{f_e^+(\theta - \tau_e), \nu_e} & \text{ if } z_e(\theta) = 0.
\end{cases}
\end{equation}

Finally, we require that the outflow rate of a commodity $j$ at time $T_e(\vartheta)$ corresponds proportionally to its
fractional part of the total inflow rate at the entrance time $\vartheta$:
 \begin{equation} \label{eq:FIFO}
f^-_{j, e} (\theta) = \begin{cases}
f^-_e (\theta) \cdot \frac{f^+_{j, e}(\vartheta)}{f^+_e(\vartheta)} & \text{ if }  f^+_e(\vartheta) > 0,\\
0 & \text{ else,}
\end{cases}
\end{equation}
where $\vartheta = \min\set{\xi\leq \theta | T_e(\xi)=\theta}$. This ensures that the proportions of the commodities are
preserved during the arc traversal. Additionally, this condition guarantees FIFO since flow of a commodity cannot
overtake flow of other commodities.

If Conditions~\eqref{eq:flow_conservation} to \eqref{eq:FIFO} are met, we call $f$ a \emph{feasible flow over time}.

\paragraph{Arrival times.} A given feasible flow over time $f$ uniquely determines the node arrival times of a
particle $\phi \in M_j$ along a simple $o_j$-$d_j$-path~$P$. Note that particle $\phi$ enters the source~$o_j$ at time
\[\l_{j, o_j}^P(\phi) \coloneqq \min \Set{ \theta  \geq 0| \int_0^\theta u_j(\xi) \diff \xi = \phi}.\]
For an arc $e = uv$ of path $P$ the \emph{arrival time} at node $v$ of particle $\phi$ is given by
\[\l_{j,v}^P(\phi) \coloneqq T_e\left(\l_{j,u}^P(\phi)\right).\]

\subsection{Multi-commodity Nash flows over time.} In order to consider dynamic equilibria in this setting, we first need
to define the earliest point in time a particle can arrive at a certain node. This depends on a given feasible flow over
time $f$ (in particular on the preceding particles), but for clearance, we omit $f$ in the following notation.

\paragraph{Earliest arrival times.} For every commodity $j \in J$ and particle $\phi \in M_j$ the earliest arrival
time of $\phi$ at some node $v$ is given by taking the minimum arrival time at node $v$ over all simple
$o_j$-$d_j$-paths~$\mathcal{P}_j$ containing $v$.  Formally,
\[\l_{j,v}(\phi)= \min_{P \in \mathcal{P}_{j} \colon v \in P} \l_{j,v}^P(\phi).\]

\paragraph{Current shortest paths network.} For every commodity $j \in J$ and every particle $\phi \in M_j$ we
define the current shortest paths network $E'_{j, \phi}$ by all arcs that are on a shortest $o_j$-$d_j$-path, i.e.,
\[E'_{j, \phi} \coloneqq \Set{e \in E | \text{ there exists a } P \in \mathcal{P}_j \text{ with } e \in P \text{ and }
\l_{j,d_j}^P(\phi) = \l_{j,d_j}(\phi)}.\]

Finally, this enables us to define a dynamic equilibrium, which is a Nash equilibrium with infinitely many players. It
is a feasible flow over time, in which (almost) every particle corresponding to a player uses a quickest
$o_j$-$d_j$-path.

\medskip
\begin{definition}
A feasible flow over time is a \emph{multi-commodity Nash flow over time} if for all arcs~$e = uv \in E$, all
commodities~$j \in J$ and almost all points in time~$\theta \in [0, \infty)$ it holds that
\[
f_{j, e}^+(\theta) > 0 \;\;\Rightarrow\;\; \theta \in \Set{\l_{j,u}(\phi) | \phi \in  M_j \text{ such that }
e \in E'_{j, \phi}}.
\]
\end{definition}
\medskip

For all networks and all supply rate functions $u_j \in L^p$ with bounded support such a Nash flow over time exists~\cite{cominetti2015existence}. 
Since this is the case in our model the existence of a Nash flow over time is always
guaranteed.

\section{Coupling particles and packets.} In the following, we connect a flow over time instance with discretization parameters to obtain a packet routing
instance. Additionally, we transfer the definitions and notations from flows over time to the packet routing model.
Overall, this section serves as a foundation for the convergence proof in \Cref{sec:convergence}.

The setting is as follows. We are given a network $G$ and a finite set of commodities $J$, each with an integrable and
bounded supply rate function $u_j: \R_{\geq 0} \to \R_{\geq 0}$ that has bounded support. Let $m_j \coloneqq
\int_0^\infty u_j(\xi) \diff \xi$ and $\bar{u}_j \coloneqq \sup_{\theta \in \R_{\geq 0}} u_j(\theta)$. As we focus on
the network loading problem we assume without loss of generality that each commodity only uses a single fixed simple
path $P_j$.

For given discretization parameters $(\alpha, \beta)$ we define for each commodity a set of discrete packets with
suitable release times and set up a toolbox of properties that connects discrete packet routings with flows over time.

\paragraph{Notation.} Let $f = (f_{j, e}^+, f_{j, e}^-)_{j \in J, e \in E}$ be a feasible multi-commodity flow over time
with functions $F_{j, e}^+$, $F_{j, e}^-$, $z_e$, $q_e$, $T_e$ and $\l_{j, v}$ as described in \Cref{sec:flowsovertime}.

\paragraph{Packets.} In order to obtain a packet flow, we consider packets $N_j \coloneqq \set{1, 2, \dots, \lfloor
m_j/\beta \rfloor}$ with path $P_j$ for every commodity $j \in J$. The release time of packet $i \in N_j$ is given by
the release time of the last particle corresponding to $i$, i.e.,
\[r_i \coloneqq \min \Set{ \theta \in \Theta_\alpha| \int_0^{\theta} u_j(\xi) \diff\xi \geq i \cdot \beta}.\]
Note that the total volume of all packets that commodity $j$ is sending is given by $\beta \cdot \abs{N_j} = \lfloor m_j
\rfloor_\beta$.

\paragraph{In- and outflow functions.} In order to describe the corresponding packet routing, the \emph{inflow
rate}~$g_{j, e}^+(\theta)$, for $\theta \in \Theta_\alpha$, denotes the combined volume of packets of commodity~$j$ that
enter arc~$e$ in time step~$\theta/\alpha$ divided by $\alpha$ (i.e., we obtain a rate in $\vol/\sec$). The
\emph{outflow rate} $g_{j, e}^-(\theta)$ is defined analogously. Formally, we have
\[ g_{j, e}^+(\theta)\coloneqq \frac{\beta}{\alpha} \cdot \abs{\set{i \in N_j | i \in L_e^+(\theta/\alpha)}} \qquad
\text{and} \qquad g_{j, e}^-(\theta) \coloneqq \frac{\beta}{\alpha} \cdot \abs{\set{i \in N_j | i \in
L_e^-(\theta/\alpha)}}.\]
The \emph{cumulative inflow} $G_{j, e}^+(\theta)$ and the \emph{cumulative outflow} $G_{j, e}^-(\theta)$ denote the
combined volume of packets of commodity $j$ that have entered/left $e$ up to time $\theta \in \Theta_\alpha$ (including
time step $t = \theta/\alpha$). They are defined as
\[G_{j,e}^+(\theta) \coloneqq \hspace{-5pt}  \sum_{\xi \in \Theta_\alpha, \xi \leq \theta} g_{j,e}^+(\xi) \cdot \alpha
\qquad \text{ and } \qquad G_{j,e}^-(\theta) \coloneqq \hspace{-5pt}  \sum_{\xi \in \Theta_\alpha, \xi \leq \theta}
g_{j,e}^-(\xi) \cdot \alpha.\]
The set of all commodities that use an arc $e$ is denoted by $J_e \coloneqq \set{j \in J|e \in P_j}$ and the set of all
commodities which pass node $v$ is denoted by $J_v \coloneqq \set{ j \in J | v \text{ lies on } P_j}$. For each arc $e$
the \emph{total inflow} function is denoted by $g_e^+ \coloneqq \sum_{j \in J_e} g_{j,e}^+$. Analogously, we define the
\emph{total outflow} function $g_e^-$ and the \emph{total cumulative flows} functions $G_e^+$ and $G_e^-$. Naturally,
all these functions map from $\Theta_\alpha$ to~$\R_{\geq 0}$. We extend this notation to $\R_{\geq 0} \to \R_{\geq 0}$
as follows.

\begin{figure}
\centering
\includegraphics{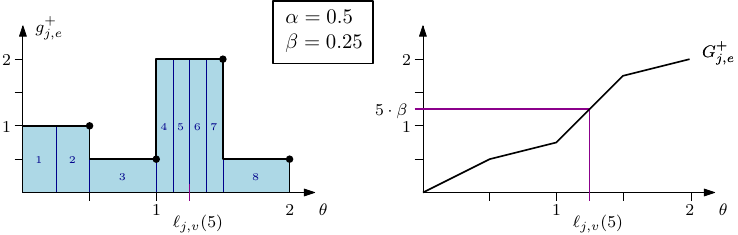}
\caption{To exemplary illustrate the refined arrival times, suppose we have $\alpha = 0.5$, $\beta = 0.25$ and eight
packets $i = 1,2,3,4,5,6,7,8$ of commodity~$j$. Suppose packet $1$ and $2$ enter arc~$e= vw$ at time $0.5$, packet $3$ at
time $1$, packet $4, 5, 6 ,7$ at time $1.5$ and packet $8$ at time $2$. As depicted on the left side the packets are
represented by a rectangle with area $\beta$ that are equally distributed under the graph of $g_{j,e}^+$ in the time
interval of length $\alpha$ right before the entrance time. The refined entrance time $\lD_{j, v}$ of packet $i$ is then
given by the right edge of the rectangle of $i$. This can also be seen by considering the cumulative inflow $G_{j, e}^+$
(right side of the figure). Hence, packet $5$ has a refined entrance time of $\lD_{j, v}(5) = 1.25$. It enters the arc
at time step $\lceil \lD_{j, v}(5) \rceil_\alpha = 1.5$ on position $k = (1.25 - (1.5 - 0.5)) / 0.5 \cdot 4 = 2$.}
\label{fig:entrance_times}
\end{figure}

\paragraph{Refined time.} In order to describe the arrival times and cumulative in- and outflows for the packet model we
use continuous piece-wise linear functions. The idea is to consider packets as a continuous flow that is equally
distributed along the time frame of one time step; see left side of \Cref{fig:entrance_times}. The in- and outflow
functions $g_{j, e}^+, g_{j, e}^-: \R_{\geq 0} \to \R_{\geq 0}$ are extended to be constant during each time step. In
other words, we extend these functions by setting $g_{j, e}^+(\theta) \coloneqq g_{j,e}^+(\lceil \theta \rceil_\alpha)$
and $g_{j,e}^-(\theta) \coloneqq g_{j,e}^-(\lceil \theta \rceil_\alpha)$ for all $\theta \in \R_{\geq 0}$. With this the
cumulative in- and outflow functions are extended by\vspace{-0.7em}
\[G_{j,e}^+(\theta) \coloneqq \int_0^\theta g_{j,e}^+(\xi) \diff \xi \qquad \text{ and } \qquad 
G_{j,e}^-(\theta) \coloneqq \int_0^\theta g_{j,e}^-(\xi) \diff \xi.\]
The \emph{refined arrival time} of packet $i \in N_j$ at node $v$ is defined as
\[\lD_{j, v}(i) \coloneqq \min \Set{ \theta \in \R_{\geq 0} | G_{j,uv}^-(\theta) \geq i \cdot \beta } 
=\min \Set{ \theta \in \R_{\geq 0} | G_{j,vw}^+(\theta) \geq i \cdot \beta }\]
for $uv, vw \in P_j$. Note that for $v = o_j$ the arrival time~$\lD_{j, v}(i)$ is defined by $G_{o_j w}^+$ with $o_jw
\in P_j$, and analogously, for $v = d_j$ it is defined by $G_{ud_j}^-$ with $ud_j\in P_j$. The superscript $D$ indicates
that we consider the discrete packet model. We also call $\lD_{j, v}(i)$ the \emph{refined entrance time} into $vw$.

As all packets of a commodity $j$ take the same path and the arc dynamics follow the FIFO principle we have that
$\lD_{j, v}(i) < \lD_{j, v}(i')$ for $i < i'$. Note that packet $i$ is processed in the node transition of $v$ at time
$\lceil \lD_{j, v}(i)\rceil_\alpha$. The refined arrival time also contains the information about the position~$k \in
\set{1, 2, \dots}$ of packet $i$ among all packets of commodity~$j$ that enter arc~$vw$ at this time. More precisely,
the position $k$ is obtained by
\[k = \frac{1}{\alpha} \cdot \big(\lD_{j, v}(i)- (\overbrace{\lceil \lD_{j, v}(i)\rceil_\alpha}^{\text{time step}}-\alpha) 
\big) \cdot \overbrace{ g_{j,vw}^+(\l_{j,v}(i)) \cdot \frac{\alpha}{\beta}}^{\text{\#packets of $j$ in $L_{vw}^+$}}.\]
A visual representation of the refined arrival times can be found in \Cref{fig:entrance_times}.

\paragraph{Queue sizes, waiting times and exit times.} We define the commodity-specific queue size~$z_{j,e}^D$, waiting
times~$q_{j,e}^D$ and exit times~$T_{j,e}^D$ for every commodity $j$ corresponding to the flow over time model.

For a commodity~$j$ the \emph{commodity-specific queue size} $z_{j,e}^D$ is given for all $\theta \in \R_{\geq 0}$ by
\[z_{j,e}^D(\theta + \tauD) \coloneqq G_{j,e}^+(\theta) - G_{j,e}^-(\theta + \tauD).\]
Note that for $\theta \in \Theta_\alpha$ the queue size $z_{j, e}^D(\theta)$ describes the total volume of packets in
$Z_e(\theta/\alpha) \cap N_j$. These are exactly the packets of commodity $j$ that have spend at least $\tauD$ time on
$e$ but are not selected to leave at time step~$\theta/\alpha$. The \emph{total queue size} (or simply \emph{queue
size}) is given by $z_e^D \coloneqq \sum_{j \in J_e} z_{j, e}^D$.
  
The \emph{waiting time} function $q_{j,e}^D$ is obtained by setting
\[q_{j,e}^D(\theta) \coloneqq \min \Set{q \geq 0 | \int_{\theta + \tauD}^{\theta + \tauD + q} g_{j,e}^-(\xi) \diff \xi 
\geq z^D_{j,e}(\theta + \tauD)}\]
for $\theta \in \R_{\geq 0}$. It describes the waiting time of packets entering arc $e$ at refined time $\theta$. With
this we obtain the \emph{exit time} function
\[T_{j,e}^D(\theta) \coloneqq \theta + \tauD + q_{j,e}^D(\theta).\]

The following lemma shows that this function indeed describes the exit time of a packet that entered $e$ at refined time $\theta$.
\begin{lemma} \label{lem:exit_times}
If packet $i \in N_j$ enters $e = uv$ at refined time $\lD_{j,u}(i) \in \R_{\geq 0}$, it leaves $e$ at refined time
\[\lD_{j,v}(i) = T_{j,e}^D(\lD_{j, u}(i)).\]
\end{lemma}
\begin{proof}
Let $\theta \coloneqq \l_{j, u}^D(i)$. By definition of $q_{j,e}^D$ together with the fact that $G_{j,e}^+(\theta) = i
\beta$ we have
\begin{align*}
T_{j,e}^D(\theta) &= \theta + \tauD + \min \Set{q \geq 0 | \int_{\theta + \tauD}^{\theta + \tauD + q} g_{j,e}^-(\xi) \diff
\xi \geq z^D_{j,e}(\theta + \tauD)}\\
&=\min \Set{T \geq \theta + \tauD | \int_{\theta + \tauD}^{T} g_{j,e}^-(\xi) \diff \xi \geq G_{j,e}^+(\theta) 
- G_{j,e}^-(\theta + \tauD)}\\
&=\min \Set{T \geq 0 | G_{j,e}^-(T)\geq G_{j,e}^+(\theta) }\\
&=\min \Set{T \geq 0 | G_{j,e}^-(T) \geq i \beta }\\
&=\lD_{j,v}(i).
\end{align*}

\vspace*{-1.75em}
\end{proof}
\medskip
 
\section{Convergence.} \label{sec:convergence}
In this section, we finally show that every feasible flow over time is the limit of the sequence of packet routings when
decreasing the packet size and time step length coherently. Observe that if $\alpha$ went quicker to zero than $\beta$,
packets would get isolated, i.e., the number of time steps between two successive packets would increase (since the
current capacity $\hat \nu_e$ would go to zero). To avoid this and, even stronger, to guarantee that the number of
packets leaving a queue per time step becomes large, we require that $\beta$ goes quicker to zero than $\alpha$. More
precisely, we only consider sequences of convergence parameters $(\alpha, \beta)$ with $\alpha \to 0$ and
$\frac{\beta}{\alpha} \to 0$ from now on.

As key component for convergence we prove the following theorem.
\begin{theorem} \label{thm:convergence}
Let $(V, E, (\nu_e)_{e \in E}, (\tau_e)_{e \in E})$ be a network and $J$ be a finite set of commodities, each $j \in J$
equipped with a simple path $P_j$ and an integrable and bounded supply rate function $u_j$ with bounded support.
Consider the packet routing induced by discretization parameters $(\alpha, \beta)$ such that $\alpha$ and
$\frac{\beta}{\alpha}$ are sufficiently small. Given a node $v$, it holds for every packet $i \in N_j$ of commodity $j
\in J_v$ and the corresponding particle $\phi\coloneqq i \beta \in M_j$ that
\[\abs{\l_{j,v}(\phi) - \l_{j,v}^D(i)} \leq \sqrt{\alpha} \cdot C^K \qquad \text{ for } K \in \N \text{ such that }
\l_{j,v}(\phi) < K \cdot \istep \text{ or } \l_{j,v}^D(i) < K \cdot \istep.\]
Here, $C > 1$ is a constant that only depends on the instance and $\istep\coloneqq \min_{e \in E} \tau_e / 2 > 0$.
\end{theorem}
To be more precise, $\alpha$ is \emph{sufficiently small} if $\sqrt{\alpha}$ is smaller than $\istep  C^{-\lceil H/\istep \rceil}$,
where $H$ is the time when the last particle left the network. For $\beta/\alpha$ to be \emph{sufficiently small} it has
to be smaller than $1$ and small enough such that there are at least two packets allowed to leave $e$ at each time step,
i.e., $\hat\nu_e = \nu_e \cdot \alpha/\beta \geq 2$ (including when entering the network, i.e., $\bar u_j \cdot \alpha/\beta \geq 2$ for all commodities $j$).

Note that this theorem implies a convergence rate of $\sqrt{\alpha}$ for the arrival times of a given
packet-particle-pair. Though, the constant might grow exponentially in the packet/particle index.

The roadmap of the proof goes as follows. We do an induction on $K$, i.e., in each induction step, we extend the validity
of the statement by $\istep$ time units. This enables us to consider each arc individually. For the induction step, we
introduce a hypothetical flow over time that matches the inflow rates of the packet model but follows the flow over time
dynamics. With this additional ingredient, we are able to show that the exit times, and hence the arrival times of the
packets and corresponding particles do not differ too much.

For an important simplification, note that packets of a commodity $j$ that enter into the network at a node $v$ behave
like packets coming from an incoming arc with capacity $\bar{u}_j$ (cf.~\Cref{alg:node_transition}). For the sake of
clarity we do not introduce additional notation but instead implicitly assume that $\delta_v^-$ includes these auxiliary
arcs with $\nu_e \coloneqq \bar{u}_j$.

Furthermore, we define the \emph{rate bound} $\kappa_e$ for all arcs $e = uv$ by
\[\kappa_e \coloneqq \max\Set{\sum_{e' \in \delta_u^-}  \left(\nu_{e'} + 1\right) ,  \nu_e + 1}.\]
This is an instance constant that only depends on the arc capacities, including the upper bounds of the supply rates.
Since the number of leaving packets of each predecessor arc $e'$ (including the auxiliary arcs for newly released
packets) is upper bounded by $\hat \nu_{e'} + 1$ in every time step, we have that $g_{e'}^-(\theta) \leq \nu_{e'} +
\beta/\alpha$. Since $\beta/\alpha$ is assumed to be smaller than $1$, $\kappa_e$ does indeed bound the maximal inflow
into $e$ as well as the maximal outflow of $e$ for both the packet model as well as the flow over time model.

We structure the proof with the help of three lemmas, which we present in the following.
The first lemma shows that the waiting time in the packet model can be approximated by the queue
size divided by the capacity.
\begin{restatable}{lemma}{lemwaitingtimeseuqlasqueuepercapa}  \label{lem:waiting_times_equals_queue_per_capa}
For all arcs $e = uv$, all commodities $j \in J_e$ and every $\theta \in \R_{\geq 0}$ the waiting time at arc $e$ in the
packet model satisfies
\[\abs{q^D_{j,e}(\theta) - \frac{ z_e^D(\theta + \tauD)}{\nu_e}} 
\leq 2 \alpha +  \frac{\alpha \kappa_e}{\nu_e}+\frac{\beta}{\nu_e}.\]
\end{restatable}
\begin{proof}
By the definition of the waiting times, we obtain for each commodity $j'$ separately that
\begin{align*}
z_{j'\!, e}^D(\theta + \tauD) &= \int_{\theta + \tauD}^{\theta + \tauD + q_{j'\!, e}^D(\theta)} g_{j'\!,e}^-(\xi) \diff \xi.
\end{align*}
Hence, since all $g_{j'\!,e}^-$ are non-negative, we obtain with $j_{\max} \coloneqq \argmax_{j' \in J} q_{j'\!,
e}^D(\theta)$ that

\begin{equation} \label{eq:discrete_queue_size}
\begin{aligned}
z_e^D(\theta + \tauD) 
&\leq \int_{\theta + \tauD}^{\theta + \tauD + q_{j_{\max},e}^D(\theta)} g_e^-(\xi) \diff \xi \\
&=  \int_{\theta + \tauD}^{\theta + \tauD + q_{j,e}^D(\theta)} g_{e}^-(\xi) \diff \xi 
+ \int_{\theta + \tauD + q_{j,e}^D(\theta)}^{\theta + \tauD + q_{j_{\max},e}^D(\theta)} g_e^-(\xi) \diff \xi.
\end{aligned}
\end{equation}
For the right term, observe that for two packets $i \in N_j$ and $i' \in N_{j'}$ entering at the same refined time (so
in particular at the same time step) the total flow volume of packets entering between them is bounded by $\alpha \cdot
\sum_{e' \in \delta_u^-} (\nu_{e'} + \beta/\alpha) \leq \alpha \kappa_e$ as this is the maximal volume that can
transition at node $u$ within one time step. As we have FIFO on the arc the same is true for cumulative outflow at the
exit times, i.e.,
\[\abs{G_e^-(T^D_{j,e}(\lD_{j,u}(i))) - G_e^-(T^D_{j'\!, e}(\lD_{j'\!, u}(i')))} \leq \alpha \kappa_e.\]
In particular, we have 
\[\int_{\theta + \tauD + q_{j,e}^D(\theta)}^{\theta + \tauD + q_{j_{\max},e}^D(\theta)} g_e^-(\xi) \diff \xi 
= G_e^-(T_{j_{\max},e}^D(\theta)) - G_e^-(T_{j,e}^D(\theta)) \leq \alpha \kappa_e.\]
For the left term of \eqref{eq:discrete_queue_size} observe that for all time steps $t$ with $\theta + \tauD < t \alpha
\leq \theta + \tauD + q_{j,e}^D(\theta) + \alpha$ the number of leaving packets $L_e^-(t)$ is bounded by the current
capacity $\hat{\nu}_e(t)$.

The number of time steps in this interval (of length $q_{j,e}^D(\theta) + \alpha$) is upper bounded by
$q_{j,e}^D(\theta) / \alpha + 2$. By the definition of the current capacity in \eqref{eq:capacity} we obtain that the
number of packets that leave $e$ within this interval is upper bounded by
\[\left(\frac{q_{j,e}^D(\theta)}{\alpha} + 2\right) \cdot \hat \nu_e + 1.\]
Here that $+1$ is due to some potential remaining capacity that is transferred to the very first time step in the interval.
Hence, we obtain with $\hat \nu_e = \frac{\alpha}{\beta} \nu_e$ that
\[\int_{\theta + \tauD }^{\theta + \tauD + q_{j,e}^D(\theta)} g_{e}^-(\xi) \diff \xi 
\leq \beta \cdot \left(\frac{q_{j,e}^D(\theta)}{\alpha} + 2\right) \cdot \hat \nu_e + \beta 
= q_{j, e}^D(\theta) \cdot \nu_e + 2 \cdot \alpha \cdot \nu_e + \beta . \]
In conclusion, \eqref{eq:discrete_queue_size} yields the following upper bound
\[z_e^D(\theta + \tauD) \leq q_{j, e}^D(\theta) \cdot \nu_e + \alpha \cdot (\kappa_e + 2 \cdot \nu_e) + \beta.\]

For the lower bound, we consider $j_{\min} \coloneqq \argmin_{j' \in J} q_{j'\!, e}^D(\theta)$, which yields
\begin{equation}\label{eq:discrete_queue_size_lower}
\begin{aligned}
z_e^D(\theta + \tauD) 
&\geq \int_{\theta + \tauD}^{\theta + \tauD + q_{j_{\min},e}^D(\theta)} g_e^-(\xi) \diff \xi \\
&=  \int_{\theta + \tauD}^{\theta + \tauD + q_{j,e}^D(\theta)} g_{e}^-(\xi) \diff \xi 
- \int_{\theta + \tauD + q_{j_{\min},e}^D(\theta)}^{\theta + \tauD + q_{j,e}^D(\theta)} g_e^-(\xi) \diff \xi.\end{aligned}
\end{equation}
With the same argument as before, we obtain for the right term that
\[\int_{\theta + \tauD + q_{j_{\min},e}^D(\theta)}^{\theta + \tauD + q_{j,e}^D(\theta)} g_e^-(\xi) \diff \xi 
\leq \alpha \cdot \kappa_e.\]
For the left term of \eqref{eq:discrete_queue_size_lower}, we consider time steps $t$ with $\theta + \tauD + \alpha \leq
t \alpha < \theta + \tauD + q_{j,e}^D(\theta)$. The number of time steps in this interval (of length $q_{j,e}^D(\theta)
- \alpha$) are bounded from below by $q_{j,e}^D(\theta) / \alpha  - 1$. Furthermore, in all these time steps the set of
leaving packets $L_e^-(t)$ has maximal size, i.e., $\abs{L_e^-(t)} = \lfloor \hat \nu_e(t) \rfloor$. This is true since
packet $i$ of commodity $j$ that has entered at $\theta$ is in $B_e(t)$ but not in $L_e^-(t)$ (since it leaves $e$ only
at time $\lceil \theta + \tauD + q_{j,e}^D(\theta) \rceil_\alpha$). Hence, with the definition of the current capacity
(see \Cref{eq:capacity}) the number of packets that leave the arc within this interval is bounded from below by
\[\left(\frac{q_{j,e}^D(\theta)}{\alpha} - 1\right) \cdot \hat \nu_e - 1.\]
Here, the last $-1$ is due to the rounding in the last time step of the interval. We obtain that
\[\int_{\theta + \tauD }^{\theta + \tauD + q_{j,e}^D(\theta)} g_{e}^-(\xi) \diff \xi 
\geq \beta \cdot \left(\frac{q_{j,e}^D(\theta)}{\alpha} - 1\right) \cdot \hat \nu_e - \beta 
= q_{j, e}^D(\theta) \cdot \nu_e - \alpha \cdot \nu_e - \beta.\]
In conclusion \eqref{eq:discrete_queue_size_lower} yields that
\[z_e^D(\theta + \tauD) \geq q_{j, e}^D(\theta) \cdot \nu_e - \alpha \cdot (\kappa_e + \nu_e) - \beta.\]
Dividing the upper and the lower bound by $\nu_e$ results in
\[ - \alpha \frac{\kappa_e}{\nu_e} - \alpha - \frac{\beta}{\nu_e} 
\leq \frac{z_e^D(\theta + \tauD)}{\nu_e} -q_{j, e}^D(\theta) 
\leq  \alpha \frac{\kappa_e}{\nu_e} + 2 \alpha + \frac{\beta}{\nu_e},\]
which completes the proof.
\end{proof}
\medskip

\paragraph{Hypothetical flow over time.} For a fixed arc $e = uv$ we introduce the \emph{hypothetical flow over time} by
setting $h_{j,e}^+ \coloneqq g_{j,e}^+$ and $h_{j,e}^- \colon \R_{\geq 0} \to \R_{\geq 0}$ according to
\eqref{eq:totaloutflow} and \eqref{eq:FIFO}. In other words, we take the inflow rates of the discrete packet model, but
determine the outflow rates according to the flow over time model. We define the cumulative flows as before by
$H_{j,e}^+(\theta) \coloneqq \int_0^{\theta} h_{j,e}^+(\xi) \diff \xi$ and $H_{j,e}^-(\theta) \coloneqq \int_0^{\theta}
h_{j,e}^-(\xi) \diff \xi$. Clearly, we have $H_{j,e}^+ = G_{j,e}^+$, but in general we cannot expect that $H_{j,e}^-$
equals $G_{j,e}^-$. Again, we define the \emph{total hypothetical cumulative flow} by $H_e^+ \coloneqq \sum_{j \in J}
H_{j,e}^+$ and $H_e^- \coloneqq \sum_{j \in J} H_{j,e}^-$. The queue size, waiting time and exit time function for the
hypothetical inflow rate are denoted by $z_e^H$, $q_e^H$ and $T_e^H$. We can even define hypothetical arrival times at
$u$ and $v$ for all $\phi \in M_j \setminus \set{0}$ by
\[k_{j, u}(\phi) \coloneqq \min \set{\theta \in \R_{\geq 0} | H_{j,e}^+(\theta) = \phi} \qquad \text{ and } \qquad 
k_{j, v}(\phi) \coloneqq \min \set{\theta \in \R_{\geq 0} | H_{j,e}^-(\theta) = \phi}.\]
With this definition, the particles $(\beta (i-1), \beta i] \subseteq M_j$ correspond to packet $i \in N_j$. Particle $0$ is a special case for which we explicitly include it into packet $1$, by defining
\[k_{j, u}(0) \coloneqq \max \set{\theta \in \R_{\geq 0} | H_{j,e}^+(\theta) = 0} \qquad \text{ and } \qquad 
k_{j, v}(0) \coloneqq \max \set{\theta \in \R_{\geq 0} | H_{j,e}^-(\theta) = 0}.\]

Clearly, we have that $k_{j, u}(i \beta) = \l_{j, u}^D(i)$ for all $i \in N_j$. Since the hypothetical flow on arc $e$
follows the continuous flow dynamics we have $k_{j, v}(\phi) = k_{j,u}(\phi) + \tau_e + q_e^H(k_{j,u}(\phi))$. Note that
the hypothetical flow rates are considered on each arc separately. In particular, we cannot expect that they satisfy
flow conservation.

In the next lemma, we show that, given some particle $\phi$, whenever the difference between the arrival times at $u$
(hypothetical flow compared to continuous flow) is not too large for all particles arriving earlier than $\phi$, then
this is also true for the difference in the arrival times of particle $\phi$ at node $v$.
\begin{restatable}{lemma}{lemboundonexittimes}   \label{lem:bound_on_exit_times}
Consider an arc $e = uv$, a commodity $j \in J_e$ and a particle $\phi \in M_j$ with $\phi \in \beta\N$. Suppose there
exists a $\delta > \alpha$ with
\[\abs{\l_{j'\!,u}(\varphi) - k_{j'\!,u}(\varphi)}\leq \delta\]
for all commodities $j' \in J_e$, and all particles $\varphi \in M_{j'}$ with $\varphi \in \beta \N$ and
$\l_{j'\!,u}(\varphi) \leq \l_{j,u}(\phi)$ or $k_{j'\!,u}(\varphi) \leq k_{j, u}(\phi)$. Then it holds that
\[\abs{\l_{j, v}(\phi) - k_{j, v}(\phi)} \leq 11  \frac{\kappa_e}{\nu_e} \cdot  \delta.\]
\end{restatable}

\begin{proof}
The proof is structured as follows. We first bound the difference in the cumulative inflow rates (between the continuous flow
and the hypothetical flow) at node $u$ at the entrance time, afterwards we bound the difference in the cumulative
outflow rates at node $v$ at the queue entrance time. Finally, we can bound the difference in the arrival time functions
at node $v$.

We start the proof with the following claim. Let $\theta \coloneqq \l_{j,u}(\phi)$ and $\zeta \coloneqq k_{j, u}(\phi)$.
\begin{claim} \label{claim:vol}
It holds that
\[\abs{F_e^+(\theta) - H_e^+(\zeta)} \leq 3  \kappa_e \cdot \delta.\]
\end{claim}
 
\begin{subproof}[Proof of \Cref{claim:vol}.]
The continuous flow and the hypothetical flow are symmetric when considering a single arc only. For that reason, we focus on the proof of $H_e^+(\theta) - F_e^+(\zeta) \leq 3  \kappa_e \cdot \delta$.

Firstly, note that $F_e^+$ and $H_e^+$ are both non-decreasing functions, whose slopes are bounded by $\kappa_e$. For
each commodity $j' \in J_e$ it holds that $F_{j'\!, e}^+(\l_{j'\!, u}(\varphi)) = \varphi = H_{j'\!,
e}^+(k_{j'\!,u}(\varphi))$ for all $\varphi \in M_{j'}$.
We partition $J_e$ into two sets:
\[J_1 \coloneqq \set{j' \in J_e | H_{j'\!, e}^+(\zeta) = 0}\qquad \text{ and } \qquad 
J_2 \coloneqq \set{j' \in J_e | H_{j'\!, e}^+(\zeta) > 0}.\]
In other words, $J_2$ is the set of commodities for which flow of the first package has already arrived at $u$ at time $\zeta$ and $J_1$ are commodities where no flow has arrived there yet.
Note that for all commodities $j' \in J_1$ it holds that $F_{j'\!,e}^+(\theta) \geq 0 = H_{j'\!,e}^+(\zeta)$.

For all $j' \in J_2$, we denote by $\phi_{j'}$ the last particle of $M_{j'}$ with $\phi_{j'} \in \beta\N$ and $k_{j'\!,
u}(\phi_{j'}) \leq \zeta$. Due to the requirements of the lemma applied to $\phi$ and $\phi_{j'}$, the arrival time of
this particle $\phi_{j'}$ in the continuous flow satisfies
\[\theta \geq \zeta - \delta \geq k_{j'\!,u}(\phi_{j'}) - \delta \geq \l_{j'\!,u}(\phi_{j'}) - 2 \delta.\]
Combining these observations yields
\begin{align*}
F_e^+(\theta) &= \sum_{j' \in J_1 } F_{j'\!,e}^+(\theta) + \sum_{j' \in J_2} F_{j'\!,e}^+(\theta)\\
& \geq \sum_{j' \in J_1} H_{j'\!,e}^+(\zeta) + \sum_{j' \in J_2} F_{j'\!,e}^+(\l_{j'\!,u}(\phi_{j'})) - 2 \delta \kappa_e\\
&= \sum_{j' \in J_1} H_{j'\!,e}^+(\zeta) +  \sum_{j' \in J_2} H_{j'\!,e}^+(k_{j'\!,u}(\phi_{j'}))  - 2 \delta \kappa_e\\
&\geq \sum_{j' \in J} H_{j'\!,e}^+(\zeta) - 2 \delta \kappa_e - \alpha \kappa_e \geq \sum_{j' \in J} H_{j'\!,e}^+(\zeta) 
- 3 \delta \kappa_e .\end{align*}
The first inequality holds since the slope of the total inflow $F_e^+$ is bounded by $\kappa_e$ and for each commodity
$j' \in J_2$ the difference $\l_{j'\!,u}(\phi_{j'}) - \theta$ is
upper bounded by $2\delta$. Finally, note that the last inequality holds since $k_{j'\!,u}(\phi_{j'})$ and $\zeta$
correspond to the same time step or no flow of commodity $j'$ enters $e$ between these points in time.

Equivalently, by replacing $\zeta$ by $\theta$, $k$ by $\ell$ and $F^+$ and by $H^+$ we obtain
\[H_e^+(\zeta) \geq F_e^+(\theta) - 3  \delta \kappa_e.\]

\vspace*{-1.75em}
\end{subproof}
\medskip
 
Since the deviation of cumulative inflow is bounded, we can derive a bound on the difference in the cumulative outflow
functions at the time the particle $\phi$ enters the respective queues.
\begin{claim} \label{claim:outflow}
We have
\[\abs{F_e^-(\theta+\tau_e)-H_e^-(\zeta+\tau_e)} \leq 7 \delta \kappa_e.\]
\end{claim}

\begin{subproof}[Proof of \Cref{claim:outflow}.]
Assume for contradiction that we have $F_e^-(\theta+\tau_e)-H_e^-(\zeta+ \tau_e) > 7 \delta  \kappa_e$.
Let $\zeta'$ be the latest point in time before $\zeta$, such that there was no queue at time $\zeta' + \tau_e$, i.e.,
\[\zeta' \coloneqq \max \set{\xi \leq \zeta | H_e^+(\xi) - H_e^-(\xi + \tau_e) = 0}.\]
Note that this condition is satisfied for $\xi = 0$, since $H_e^+(0) = H_e^-(\tau_e) = 0$. Since either $\zeta' = \zeta$
or a queue builds up immediately after $\zeta'$ (implying $H_e^+$ is strictly increasing), there has to be a commodity
$j'$ and a particle $\phi' \in \beta\N$ with $k_{j'\!,u}(\phi') \in [\zeta', \zeta' + \alpha]$. Note that this
immediately implies that
\[H_e^+(k_{j'\!,u}(\phi')) - H_e^+(\zeta') \leq \alpha \kappa_e \leq \delta \kappa_e.\]
Since there is no queue for $\zeta'$, it holds that $H_e^-(\zeta' + \tau_e)= H_e^+(\zeta')$. Furthermore, within
$(\zeta' + \tau_e, \zeta + \tau_e]$ there is a queue present on $e$, and thus, we have $h_e^-(\xi) = \nu_e$ within this
interval. Hence,
\[H_e^-(\zeta + \tau_e ) = H_e^-(\zeta' + \tau_e) + \nu_e \cdot (\zeta - \zeta').\]
To proceed, we distinguish two cases. 
\paragraph{Case 1:} $\theta \geq \zeta' +  \delta$.\\ 
Since queues can never be negative, ($F_e^+(\vartheta) - F_e^-(\vartheta +\tau_e) \geq 0$ for all $\vartheta$) and $\zeta
\in [\theta-\delta, \theta+\delta]$ we obtain
\[F_e^+( \zeta'+\delta)  \geq F_e^-(\zeta' + \delta+\tau_e) 
\geq F_e^-(\theta + \tau_e) - \nu_e \cdot (\theta -(\zeta'+\delta)) 
\geq F_e^-(\theta + \tau_e) - \nu_e \cdot (\zeta - \zeta').\]
Note for the second inequality that the slope of $F_e^-$ is bounded by $\nu_e$.
By the requirement of the lemma, it follows that $\l_{j'\!,u}(\phi') \in [\zeta'-\delta, \zeta'+2\delta]$, and since
$F_e^+$ cannot grow faster than $\kappa_e$ we obtain
\[\abs{F_e^+(\zeta'+\delta) - F_e^+(\l_{j'\!,u}(\phi'))} \leq 2 \delta \kappa_e.\]
Putting all together, we get
\begin{align*}
F_e^+(\l_{j'\!,u}(\phi')) + 3 \delta \kappa_e - H_e^+(k_{j'\!,u}(\phi'))
&\geq F_e^+(\zeta'+\delta)-H_e^+(\zeta') \\
&\geq F_e^-(\theta + \tau_e) - \nu_e \cdot (\zeta - \zeta') - H_e^-(\zeta' + \tau_e) \\
&= F_e^-(\theta + \tau_e) - \nu_e \cdot (\zeta - \zeta') - (H_e^-(\zeta + \tau_e) - \nu_e \cdot (\zeta - \zeta'))\\
&= F_e^-(\theta + \tau_e) - H_e^-(\zeta + \tau_e) \\
&> 7 \delta \kappa_e.
\end{align*}
This is a contradiction to \Cref{claim:vol} (applied to particle $\phi'$ of commodity $j'$, which satisfies the requirement of \Cref{lem:bound_on_exit_times}).

\paragraph{Case 2:} $\theta < \zeta' +  \delta$.\\
By the requirement of the lemma, we have $\zeta' \leq \zeta \leq \theta + \delta$, which gives us $\abs{\theta-
\zeta'}\leq\delta$.
Using this and the fact that queues cannot be negative yields
\[F_e^+(\zeta') \geq F_e^-(\zeta'+\tau_e) \geq F_e^-(\theta - \delta +\tau_e) 
\geq F_e^-(\theta + \tau_e)-\nu_e \delta.\]
By the requirement of the lemma, it follows that $\l_{j'\!,u}(\phi') \in [\zeta'-\delta, \zeta'+2\delta]$, and since
$F_e^+$ cannot grow faster than $\kappa_e$, we obtain
\[\abs{F_e^+(\zeta') - F_e^+(\l_{j'\!,u}(\phi'))} \leq 2\delta \kappa_e.\]
Again, putting everything together gives
\begin{align*}
F_e^+(\l_{j'\!,u}(\phi')) + 3\delta \kappa_e - H_e^+(k_{j'\!,u}(\phi')) &\geq
F_e^+(\zeta')-H_e^+(\zeta') \\
&= F_e^+(\zeta')  - H_e^-(\zeta' + \tau_e) \\
&\geq F_e^-(\theta + \tau_e)- \nu_e \delta - H_e^-(\zeta + \tau_e) \\
&> 7   \delta \kappa_e - \nu_e \delta\\
&\geq 6 \delta \kappa_e.
\end{align*}
This again contradicts \Cref{claim:vol}. 
Hence, in both cases, we obtain 
\[F_e^-(\theta+\tau_e)-H_e^-(\zeta+ \tau_e) \leq 7 \delta \kappa_e.\]
The assumption $H_e^-(k_{j,u}(\phi)+ \tau_e)- F_e^-(\l_{j,u}(\phi)+\tau_e) > 7  \delta \kappa_e$ can lead to a
contradiction in the same way by using the symmetry of flow $F$ and $H$. 
\end{subproof}
\medskip

We can now use \Cref{claim:vol} and \Cref{claim:outflow} to finish the proof of the lemma.
The difference in the arrival times of particle $\phi \in \beta\N$ at node $v$ in the continuous flow and hypothetical
flow can be bounded as follows:
\begin{align*}
\abs{\l_{j, v}(\phi) - k_{j, v}(\phi)}&=  \abs{\l_{j, u}(\phi) + \tau_e + q_{j,e}(\l_{j,u}(\phi)) 
- (k_{j, u}(\phi)+ \tau_e +q^H_{e,j}(k_{j, u}(\phi))) }\\
&\leq \delta + \abs{q_{j,e}(\l_{j,u}(\phi))-q^H_{e,j}(k_{j, u}(\phi))}\\
&= \delta + \abs{\frac{1}{\nu_e} \cdot \left(F_e^+(\l_{j,u}(\phi))- F_e^-(\l_{j,u}(\phi)+\tau_e)-H_e^+(k_{j, u}(\phi)) 
+ H_e^-(k_{j, u}(\phi)+\tau_e)\right) }\\
&\leq \delta + \frac{1}{\nu_e}\left( 3 \delta \kappa_e + 7  \delta \kappa_e\right) \\
&\leq  11 \delta \frac{\kappa_e}{\nu_e}. 
\end{align*}

\vspace*{-1.75em}
\end{proof}
\medskip

In the next lemma we bound the difference between the hypothetical arrival time $k_{j, v}$ and the refined packet
arrival time $\l_{j, v}^D$.
\begin{restatable}{lemma}{lemarcroundingerror} \label{lem:arc_rounding_error}
Given an arc $e = uv$, a point in time $\theta$ and a commodity $j \in J_e$. For sufficiently small $\alpha$ it holds
that
\[\abs{k_{j, v}(\phi) - \l_{j, v}^D(\lceil{\textstyle \frac{\phi}{\beta}}\rceil)} 
\leq C_1 \cdot \sqrt{\alpha} + \theta \cdot C_2 \cdot \sqrt{\alpha} \qquad 
\text{ for } \phi \in M_j \text{ with } k_{j, u}(\phi) \leq \theta.\]
Here, $C_1$ and $C_2$ are positive constants that only depend on the 
instance properties $\kappa_e$ and $\nu_e$.
\end{restatable}


\begin{proof}
This lemma bounds the error between the arrival times in the packet model and the continuous model on a single arc. To
do so, we introduced the hypothetical flow, which has exactly the same inflow functions as the packet routing. In the
following, we bound the difference between the arrival time of a packet and the corresponding particle entering arc $e$ at the
same time.
 
As we only consider a single arc, on which we have FIFO in both models, we suppose for the beginning that all particles
and packets belong to a single commodity with particles $\phi \in M \coloneqq [0, m] \subset \R_{\geq 0}$, with $m
\coloneqq \sum_{j \in J} m_j$, and packets $i \in N \coloneqq \Set{1, 2, \dots, \sum_{j \in J} \abs{N_j}}$,
respectively. In other words, we focus on the total flow.
We define the arrival times at $u$ and $v$ by
\begin{align*}
k_u(\phi) &\coloneqq \min \set{\theta \in \R_{\geq 0} | H^+_e(\theta) = \phi}, \qquad &k_v(\phi) 
&\coloneqq \min \set{\theta \in \R_{\geq 0} | H^-_e(\theta) = \phi},\\
\l_u^D(i)&\coloneqq \min \set{\theta \in \R_{\geq 0}  | G^+_e(\theta) = i \beta}, \qquad &\l_v^D(i) 
&\coloneqq \min \set{\theta \in \R_{\geq 0} | G^-_e(\theta) = i \beta}.
\end{align*}
Clearly, as we have $H_e^+ = G_e^+$, it holds that $k_u(i \beta) = \l_u^D(i)$ for all $i \in N$.

From now on, we consider a particle-packet-pair $\phi$ and $i \coloneqq \lceil \frac{\phi}{\beta} \rceil$. We will show
by induction over the particles that the following holds.
\begin{claim}
\label{claim:hypothetical_discrete_error} There exists constants $\tilde{C}_1$ and $\tilde{C}_2$ that only depend on the
instance properties such that for all $\phi \in M$ with $i = \lceil \frac{\phi}{\beta}\rceil \in N$ it holds that
\[\abs{k_v(\phi) - \l_v^D(i)} \leq \delta(\phi) \qquad \text{ with } \quad \delta(\phi) 
\leq \tilde{C}_1 \cdot \sqrt{\alpha} + \phi \cdot \tilde{C}_2 \cdot \sqrt{\alpha}.\]
\end{claim}

\begin{subproof}[Proof of \Cref{claim:hypothetical_discrete_error}.]
The proof idea is to do an induction over the particles $M \subseteq \R_{\geq 0}$. In every induction step we extend the
statement for a flow volume of at least $\sqrt{\alpha} - \beta$. The major part of the proof deals with the induction
step and the induction start is shown at the very end.

Through the proof, we need some particular particles and packets which we denote as follows.
Let $\phi'$ be the particle that leaves $e$ at the moment when particle $\phi$ enters the queue. Formally, $\phi'$ is
unique particle with $k_v(\phi') = k_u(\phi)+\tau_e$. Note that whenever $z_e^H(k_u(\phi) + \tau_e) > 0$ there has to be
a particle leaving the queue of $e$ at that very moment. Otherwise $\phi' = \phi$. Let $i'$ be the particle
corresponding to $\phi'$, formally $i'\coloneqq \lceil \frac{\phi'}{\beta}\rceil$.

Let furthermore $j$ be the last packet that left the queue before packet $i$ lines up in the packet routing, in other
words, $j \coloneqq \max{\set{j' \leq i | \l_v^D(j') \leq \l_u^D(i) +\tauD}}$. This set is not empty as it contains
packet $1$. We denote the corresponding particle by $\psi \coloneqq j \cdot \beta$.
Overall, we have the following correspondences between packets and particles:
\[i \leftrightarrow \phi \qquad i' \leftrightarrow \phi' \qquad j \leftrightarrow \psi.\]
In order to bound the difference between $k_v(\phi)$ and $\l_v^D(i)$, we carefully analyze the arrival times of these
particles and packets.

First of all, we assume that $\frac{\beta}{\alpha}$ is small enough such that there is at least one packet allowed to
leave $e$ at each time step, i.e., $\hat \nu_e \geq 1$. This way, we ensure that
\begin{equation} \label{eq:diff_i_queue_entering_j_arriving}
\abs{\l_u^D(i) + \tauD - \l_v^D(j)} \leq \alpha.
\end{equation}
Furthermore, since $G_e^+=H_e^+$, we have that $\abs{k_u(\phi) - \l_u^D(i)} \leq \alpha$, and hence,
\begin{equation} \label{eq:two_alpha_induction}
\abs{k_v(\phi')-\l_v^D(j)} \stackrel{\eqref{eq:diff_i_queue_entering_j_arriving}}{\leq} \abs{k_u(\phi) 
+ \tau_e -(\l_u^D(i) + \tauD)} + \alpha  \leq  3 \alpha.
\end{equation}
In the following, we distinguish three cases in dependency of the queue lengths at the time particle $\phi$ or packet
$i$, respectively, queues up:
\begin{enumerate}
\item The continuous and discrete queue are small: $z^H_e(k_u(\phi)+\tau_e) \leq \sqrt{\alpha}$ and
$z^D_e(\l_u^D(i)+\tauD)\leq \sqrt{\alpha}$.
\item The continuous queue is larger than $\sqrt{\alpha}$ and larger than the discrete queue: 
$z^H_e(k_u(\phi)+\tau_e) > \sqrt{\alpha}$ and $z^D_e(\l_u^D(i)+\tauD)\leq z^H_e(k_u(\phi)+\tau_e)$.
\item The discrete queue is larger than $\sqrt{\alpha}$ and larger than the continuous queue: 
$z^D_e(\l_u^D(i)+\tauD) > \sqrt{\alpha}$ and $z^D_e(\l_u^D(i)+\tauD) > z^H_e(k_u(\phi)+\tau_e)$.
\end{enumerate}

\paragraph{Case 1:} $z^H_e(k_u(\phi)+\tau_e) \leq \sqrt{\alpha}$ and $z^D_e(\l_u^D(i)+\tauD)\leq \sqrt{\alpha}$.\\ Since
both queues are upper bounded by $\sqrt{\alpha}$, we obtain
\[k_v(\phi) \in \left[k_v(\phi'), k_v(\phi')+\frac{\sqrt{\alpha}}{\nu_e}\right].\]
With \Cref{lem:waiting_times_equals_queue_per_capa} and \eqref{eq:diff_i_queue_entering_j_arriving}, we obtain
\[\l_v^D(i) \in \left[\l_v^D(j), \l_v^D(j) + \frac{\sqrt{\alpha}}{\nu_e} + 3\alpha + \frac{\kappa_e \alpha}{\nu_e} 
+ \frac{\beta}{\nu_e}\right].\]
If we put this together with \eqref{eq:two_alpha_induction}, it follows that:
\[\abs{k_v(\phi)-\l_v^D(i)} \leq \frac{\sqrt{\alpha}}{\nu_e}+6 \alpha+\frac{\kappa_e \alpha}{\nu_e}+\frac{\beta}{\nu_e}.\]

\paragraph{Case 2:} $z^H_e(k_u(\phi)+\tau_e) > \sqrt{\alpha}$ and $z^D_e(\l_u^D(i)+\tauD)\leq
z^H_e(k_u(\phi)+\tau_e)$.\\ For the sake of simplicity, we use the intuitive notation and write $x \pm y$ for the
interval $[x-y, x+y]$.
As a first step, notice that
\[ k_v(\phi) = k_v(\phi')+\frac{(\phi-\phi')}{\nu_e} \in k_v(\phi')+\frac{\beta( i - i')}{\nu_e} \pm \frac{\beta}{\nu_e}\]
and similar with \Cref{lem:waiting_times_equals_queue_per_capa}
\begin{align*}
\l_v^D(i)  & \stackrel{\eqref{eq:diff_i_queue_entering_j_arriving}}{\in} \l_v^D(j) + q_e^D(\l_u^D(i)) \pm \alpha
\subseteq \l_v^D(j)+\frac{\beta (i-j)}{\nu_e} \pm \left(3 \alpha +\frac{\kappa_e \alpha}{\nu_e} + \frac{\beta}{\nu_e}\right).
\end{align*}
This gives us that
\begin{equation} \label{eq:this_gives_us}
\begin{aligned}
\abs{k_v(\phi)-\l_v^D(i)} &\leq \abs{k_v(\phi')-\l_v^D(j) + \frac{\beta ((i - i')- (i-j))}{\nu_e} } + 3 \alpha
+\frac{\kappa_e \alpha}{\nu_e}+  \frac{2\beta}{\nu_e}\\
&=  \abs{k_v(\phi')-\l_v^D(j) + \frac{\beta(j - i')}{\nu_e}} +  3 \alpha +\frac{\kappa_e \alpha}{\nu_e}
+\frac{2\beta}{\nu_e}.
\end{aligned}
\end{equation}
Furthermore, in the case of $j\geq i'$, we observe that the difference between the arrival times at node~$v$ from packet $j$
and $i'$ can be bounded from \textbf{below} by the time the packets between them necessarily need to leave the arc $e$.
Hence, we apply \Cref{lem:waiting_times_equals_queue_per_capa} on a fictional packet routing in which packet $i'$ leaves
the queue exactly at the time packet $j$ enters the queue (consisting of all particles between $i'$ and $j$). As the
time between $i'$ and $j$ arriving at $v$ in the actual packet routing is at least that large, we obtain
\begin{equation}\label{eq:hypothetical_queue}
\l_v^D(j) - \l_v^D(i') \geq \abs{ \frac{\beta(j-i')}{\nu_e} } -\left( 2 \alpha + \frac{\kappa_e \alpha }{\nu_e}
+\frac{\beta}{\nu_e}\right). 
\end{equation}
In the case that $i' > j$, we have $ i' \leq j+2$, as we explain in the following. We have
\[(i'-1) \beta \leq \phi' = \phi - z^H_e(k_u(\phi) + \tau_e) \leq i \beta - z^H_e(k_u(\phi) + \tau_e)\]
and also
\[z_e^D(\l_u^D(i) + \tauD) = G_e^+(\l_u^D(i)) - G_e^-(\l_u^D(i) + \tauD) \geq i \beta - (j+1) \beta .\]
Hence, we obtain with the assumption of Case 2 that
\begin{align*}
(i' - j) \beta &\leq  i \beta - z^H_e(k_u(\phi) + \tau_e) +\beta - (i \beta - z_e^D(\l_u^D(i) + \tauD)- \beta) \\
&= z_e^D(\l_u^D(i) + \tauD) - z^H_e(k_u(\phi) + \tau_e) + 2\beta \\
&\leq 2\beta.
\end{align*}
In other words, there is at most one additional packet between $i'$ and $j$. Furthermore, it holds that $i' \in [j + 1, i]$, and thus,
$i'$ is in a queue at time $\l_u^D(i) + \tauD$. Assuming that $\hat \nu_e \geq 2$ (true for sufficiently small
$\frac{\beta}{\alpha}$) yields $\l_v^D(j) - \l_v^D(i') \geq - \alpha$. With $\abs{\frac{(j - i')\beta}{\nu_e}} \leq
\frac{2 \beta}{\nu_e} \leq \frac{\alpha \kappa_e}{\nu_e} + \frac{\beta}{\nu_e}$ \Cref{eq:hypothetical_queue} also holds
in this case of $i' > j$.
Finally, combining all observations above, we obtain
\begin{align*}
&\abs{k_v(\phi)-\l_v^D(i)} \\
&\stackrel{\eqref{eq:this_gives_us}}{\leq} \max\Set{ \l_v^D(j) -k_v(\phi') +  \frac{2\beta}{\nu_e}, k_v(\phi')- \l_v^D(j) 
+\abs{\frac{\beta(j - i')}{\nu_e}}} + 3 \alpha +\frac{\kappa_e \alpha}{\nu_e}+ \frac{2\beta}{\nu_e}\\
&\stackrel{\eqref{eq:hypothetical_queue}}{\leq} \max\Set{ \l_v^D(j) -k_v(\phi') +  \frac{2\beta}{\nu_e},k_v(\phi') 
-\l_v^D(i') +  2 \alpha + \frac{\alpha \kappa_e}{\nu_e}+  \frac{\beta}{\nu_e}} +3 \alpha +\frac{\kappa_e \alpha}{\nu_e}
+\frac{2\beta}{\nu_e}\\
&\stackrel{\eqref{eq:two_alpha_induction}}{\leq} \max\Set{  3 \alpha,  k_v(\phi') - \l_v^D(i') +  2 \alpha 
+\frac{\alpha \kappa_e}{\nu_e}}  +3 \alpha +\frac{\kappa_e \alpha}{\nu_e}+ \frac{4\beta}{\nu_e}\\
& \leq \delta(\phi - \sqrt{\alpha}) +6 \alpha +\frac{2\kappa_e \alpha}{\nu_e}+ \frac{4\beta}{\nu_e}
\end{align*}
Here, the last inequality follows by induction since $\phi-\sqrt{\alpha} \geq \phi'$ (case assumption) and by
monotonicity of $\delta$.

\paragraph{Case 3:} $z^D_e(\l_u^D(i)+\tauD) > \sqrt{\alpha}$ and $z^D_e(\l_u^D(i)+\tauD) > z^H_e(k_u(\phi)+\tau_e)$.\\
We begin with \[ k_v(\phi) = k_v(\phi')+\frac{(\phi-\phi')}{\nu_e}.\]
\Cref{lem:waiting_times_equals_queue_per_capa} and \eqref{eq:diff_i_queue_entering_j_arriving} provides
\begin{align*}
\l_v^D(i) & = \l_u^D(i) + \tauD + q_e^D(\l_u^D(i)) \\
&\in \l_v^D(j) +\frac{z_e^D(\l_u^D(i) + \tauD)}{\nu_e} \pm \left(3 \alpha +\frac{ \kappa_e\alpha}{\nu_e}
+\frac{\beta}{\nu_e}\right)\\
&\subseteq \l_v^D(j) +\beta \cdot \frac{i-j}{\nu_e} \pm \left(3 \alpha +\frac{\kappa_e \alpha}{\nu_e}
+\frac{2\beta}{\nu_e}\right)\\
&\subseteq \l_v^D(j)+ \frac{\phi-\psi}{\nu_e} \pm \left(3 \alpha +\frac{\kappa_e \alpha}{\nu_e}+\frac{3 \beta}{\nu_e}\right).
\end{align*}
This gives us that 
\begin{align}\label{eq:this_gives_us_case3}
\abs{\l_v^D(i) - k_v(\phi)} &\leq \abs{\l_v^D(j) - k_v(\phi') + \frac{(\phi - \psi) - (\phi-\phi')}{\nu_e}}  
+\left(3 \alpha +\frac{\kappa_e \alpha}{\nu_e}+\frac{3 \beta}{\nu_e}\right)
\end{align}
The case assumption states that the discrete queue is longer than the hypothetical queue. Together with $\phi' = \phi -
z^H_e(k_u(\phi) + \tau_e)$ and $\psi = j \beta \leq i \beta  - z_e^D(\l_u^D(i) + \tauD)$ this implies
\begin{equation} \label{eq:phi_prime_minus_psi_bigger_minus_beta}
\phi' - \psi \geq z_e^D(\l_u^D(i) + \tauD) - z^H_e(k_u(\phi) + \tau_e) + \phi -i \beta > -\beta.
\end{equation}
Furthermore, since queues in the continuous flow model cannot operate faster than $\nu_e$, we have for $\phi' \geq \psi$ that
\begin{equation}\label{eq:hypothetical_queue_case3}
k_v(\phi') -k_v(\psi) \geq  \frac{\phi'-\psi}{\nu_e}
\end{equation}
For $\phi' < \psi$ there is a queue for all particles in $[\phi, \phi'] \ni \psi$. Thus,
\eqref{eq:hypothetical_queue_case3} holds with equality.
Combining these inequalities, we obtain
\begin{align*}
&\abs{  \l_v^D(i) - k_v(\phi)}\\
&\hspace*{-2.6mm}\stackrel{\eqref{eq:this_gives_us_case3},
\eqref{eq:phi_prime_minus_psi_bigger_minus_beta}}{\leq}\hspace*{-1.8mm} \max \Set{k_v(\phi') - \l_v^D(j) 
+ \frac{\beta}{\nu_e}, \l_v^D(j) - k_v(\phi')  + \frac{(\phi' - \psi)}{\nu_e} }+ 3 \alpha 
+\frac{\kappa_e \alpha}{\nu_e}+\frac{3 \beta}{\nu_e} \\
&\stackrel{\eqref{eq:hypothetical_queue_case3}}{\leq} \max \Set{k_v(\phi') - \l_v^D(j) + \frac{\beta}{\nu_e}, \l_v^D(j) 
- k_v(\phi') + k_v(\phi') - k_v(\psi) } + 3 \alpha +\frac{\kappa_e \alpha}{\nu_e}+\frac{3 \beta}{\nu_e} \\
&\stackrel{\eqref{eq:two_alpha_induction}}{\leq}  \max\Set{ 3\alpha + \frac{\beta}{\nu_e}, \l_v^D(j)  - k_v(\psi)} 
+ 3 \alpha +\frac{\kappa_e \alpha}{\nu_e}+\frac{3 \beta}{\nu_e} \\
&\hspace*{.7mm}\leq \delta(\phi - \sqrt{\alpha}+ \beta) + 6 \alpha +\frac{\kappa_e \alpha}{\nu_e}+\frac{4 \beta}{\nu_e}.
\end{align*}
The last inequality holds by monotonicity of $\delta(\cdot)$ and since we have $\psi=j \beta  \leq i \beta -
\sqrt{\alpha} \leq \phi+ \beta -\sqrt{\alpha}$, which holds by the case assumption as the discrete queue is longer than
$\sqrt{\alpha}$.

\paragraph{Case conclusion.} To conclude the case distinction and prove the induction step, we need a monotonically
increasing function $\delta \colon M \to \R_{>0}$ that bounds $\abs{\l_v^D(i) - k_v(\phi)}$, i.e., satisfies:
\begin{itemize}
\item Case 1: $\delta(\phi) \geq 6 \alpha+\frac{\sqrt{\alpha}}{\nu_e}+\frac{\kappa_e \alpha}{\nu_e}+\frac{\beta}{\nu_e}$
\item Case 2: $\delta(\phi) \geq  \delta(\phi - \sqrt{\alpha}) +6 \alpha +\frac{2\kappa_e \alpha}{\nu_e}
+\frac{4\beta}{\nu_e}$
\item Case 3: $\delta(\phi) \geq \delta(\phi - \sqrt{\alpha}+ \beta) + 6 \alpha +\frac{\kappa_e \alpha}{\nu_e}
+\frac{4 \beta}{\nu_e}$
\item  $\delta(\phi) \to 0$ for $\alpha \to 0$ and $\frac{\beta}{\alpha} \to 0$ for all $\phi \in M$.
\end{itemize}
The function
\[\delta(\phi)= 6 \alpha+\frac{\sqrt{\alpha}}{\nu_e}+\frac{\kappa_e \alpha}{\nu_e}+\frac{2\beta}{\nu_e} 
+ \frac{2 \phi}{\sqrt{\alpha}} \left(6 \alpha +\frac{2\kappa_e \alpha}{\nu_e}+ \frac{4\beta}{\nu_e}\right) 
\leq \tilde{C}_1 \cdot \sqrt{\alpha} + \phi \cdot \tilde{C}_2 \cdot \sqrt{\alpha},\]
fulfills all requirements, where $\tilde{C}_1$ and $\tilde{C}_2$ are constants that only depend on instance properties, namely $\nu_e$
and $\kappa_e$.

It remains to show the base case, which consists of all particles in $\phi \in [0, \sqrt{\alpha}]$.
In this case, the total flow volume of packets on the arc, and hence, in the queue, is upper bounded by $\sqrt{\alpha} +
\beta$ (note that for $\phi = \sqrt{\alpha}$ the packet $i$ might be rounded up to at most $\sqrt{\alpha}/\beta + 1$).
Applying the argumentation used in Case 1 to the situation where queues are bounded by $\sqrt{\alpha}+\beta$ in both
models yields
\[\abs{  \l_v^D(i) - k_v(\phi)} \leq 6 \alpha+\frac{\sqrt{\alpha}}{\nu_e}+\frac{\kappa_e \alpha}{\nu_e}
+\frac{2\beta}{\nu_e} \leq \delta(\phi).\] 
Our choice of $\delta$ satisfies this.

This finishes the proof of the claim. Thus, we have shown that, in a single commodity setting, the difference in the
arrival times at node $v$ between a particle and the corresponding packet is upper bounded by $\delta$ which goes to $0$ when
the discretization parameters go to $0$. 
\end{subproof}
\medskip

To prove the lemma two things remain to show. First, we reassign the commodities to the packets, and second, we switch
from a particle-dependent bound to a time-dependent bound.

Given a particle $\phi_j \in M_j$ and corresponding packet $i_j \coloneqq \lceil\frac{\phi_j}{\beta}\rceil \in N_j$, let
$\vartheta \coloneqq k_{j,u}(\phi_j)$ be the point in time the particle enters $e$. We consider the same particle in the
single commodity flow setting and denote it by $\phi$. In other words, $\phi$ is the particle with $k_u(\phi) = \vartheta$.
The corresponding packet is $i \coloneqq \lceil\frac{\phi}{\beta}\rceil$. Since the flow dynamics and the inflow rates
of the single-commodity hypothetical flow equal the multi-commodity hypothetical flow, we have
\[k_v(\phi)= T^H_e(k_u(\phi)) = T^H_e(\vartheta) = T^H_e(k_{j,u}(\phi_j)) = k_{j,v}(\phi_j).\]

In the following, we argue that the arrival times at node $v$ of packet $i$ and $i_j$ are not too far from each other.
First, observe that packet $i_j$ and $i$ enter $e$ at the same time step, which is given by $\lceil \vartheta
\rceil_\alpha$. Though, the refined entrance times (which are commodity-dependent) might differ up to $\alpha$, which
might cause a small error for the exit times. The packet volume which enters between the refined entrance times of these
two packets is bounded by $\kappa_e \alpha$. Since packet $i$ and $i_j$ enter the arc at the same time step, they are
also assigned to $Q_e$ at the same time step. This implies that in the time steps between the exit times of the two
packets the capacity of the arc is fully used, i.e., at least $\lfloor \nu_e \rfloor_{\beta}$ volume leaves per time
step. Hence, the time steps when $i$ and $i_j$ leave the arc cannot be further away than $\frac{\kappa_e \alpha}{\lfloor
\nu_e \rfloor_{\beta}} + \alpha$. Note that the additive $\alpha$ is due to the rounding to time steps. For the refined
times, there might be an additional deviation of $\alpha$, which gives us
\[\abs{\l_{j,v}^D(i_j) - \l_v^D(i)}\leq  \frac{\kappa_e \alpha}{\lfloor \nu_e \rfloor_{\beta}} + 2\alpha.\]
These considerations together with \Cref{claim:hypothetical_discrete_error} lead to
\begin{align*}
\abs{\l_{j,v}^D(i_j) - k_{j,v}(\phi_j)} &\leq \abs{\l_{j,v}^D(i_j) -\l_v^D(i)} + \abs{\l_v^D(i) - k_v(\phi)} 
+\abs{ k_v(\phi)-k_{j,v}(\phi_j)}\\
&\leq \frac{\kappa_e \alpha}{\lfloor \nu_e \rfloor_{\beta}} + 2\alpha + \tilde{C}_1 \cdot \sqrt{\alpha} 
+\phi \cdot \tilde{C}_2 \cdot \sqrt{\alpha}\\
&\leq C_1 \cdot \sqrt{\alpha} + \phi \cdot \tilde{C}_2 \cdot \sqrt{\alpha}
\end{align*}
with $C_1 \coloneqq \frac{2\kappa_e}{\nu_e} + 2 + \tilde{C}_1$. Note that the last inequality only holds for
sufficiently small $\beta$, which fulfills $\lfloor \nu_e \rfloor_\beta \geq \nu_e / 2$.
In order to switch from particle $\phi$ to entrance time $\vartheta$, we use the fact that until time $\vartheta$ at most
a total flow of $\kappa_e \cdot \vartheta$ entered arc~$e$.
By choosing $C_2 \coloneqq \kappa_e \cdot \tilde{C}_2$, we obtain
\begin{align*}
\abs{\l_{j,v}^D(i_j) - k_{j,v}(\phi_j)} &\leq C_1 \cdot \sqrt{\alpha} + \phi \cdot \tilde{C}_2 \cdot \sqrt{\alpha}\\
 &\leq C_1 \cdot \sqrt{\alpha} + \vartheta \cdot \kappa_e \cdot  \frac{C_2}{\kappa_e} \cdot \sqrt{\alpha}\\
 &=C_1 \cdot \sqrt{\alpha} + \vartheta  \cdot C_2 \cdot \sqrt{\alpha}. 
\end{align*}
Since $\vartheta = k_{j,u}(\phi_j) \leq \theta$, this completes the proof of the lemma.
\end{proof}
\medskip

With the help of the previous lemmas, we can finally prove \Cref{thm:convergence}:

\begin{proof}[Proof of \Cref{thm:convergence}.]
We prove this theorem by an induction on $K$. Due to the finitely lasting network inflow rates, there is a finite time horizon $H$, for which the last particle has left the network. It is sufficient to prove the statement only for $K \in \set{1, 2, \dots, \hat H}$ with $\hat H \coloneqq \lceil H/\istep \rceil$, as they cover all arrival times at all nodes for all particles. 
For larger values of $K$ the right-hand side of the bound only grows since $C>1$, so the statement remains valid.

We choose $\alpha$ and $\beta$ to be small enough, such that $\sqrt{\alpha} < \istep C^{-\hat H}$ and $\beta/\alpha < 1$ as well as $\nu_e \cdot \alpha / \beta \geq 2$ for all $e \in E$. (The last two conditions are needed for \Cref{lem:bound_on_exit_times}.)

For the base case, we consider a point in time $\theta \in [0, \istep)$. In
this case, we only have to consider the nodes $o_j$ for each commodity $j$ since for all other nodes there is no particle
$\phi \in M_j$ with $\l_{j, v}(\phi) \leq \theta$ and also no packet $i$ with $\lD_{j, v}(i) \leq \theta$.
For the release times, we get
\begin{align*}
\lD_{j, o_j}(i)  = r_i &= \min \Set{ \theta \in \Theta_\alpha| \int_0^{\theta} u_j(\xi) \diff\xi \geq i \beta  }\\
&= \min \Set{ \theta \in \Theta_\alpha| \int_0^{\theta} u_j(\xi) \diff\xi \geq \phi}
\in [\l_{j,{o_j}}(\phi), \l_{j,{o_j}}(\phi) + \alpha].
\end{align*}
 
For the induction step to $K \leq \hat H$, assume the theorem is proven for $K-1$. Consider a packet $i \in N_j$ and a particle $\phi
\coloneqq i \beta \in M_j$ with  $\min\! \set{\!\l_{j,v}(\phi), \lD_{j,v}(i)\!} < K \istep$. If $i$ is released at node
$v$, the statement holds with the considerations of the base case. Otherwise, for the node $u$ with $e = uv \in P_j$ it
holds that $\min\! \set{\!\l_{j,u}(\phi),\lD_{j,u}(i)\!} \leq \min\! \set{\!\l_{j,v}(\phi),\lD_{j,v}(i)\!} - \tau_e \leq
(K - 1) \istep$, and therefore, we can apply the induction hypothesis to node $u$ to obtain
\begin{equation} \label{eq:induction_step_max}
\max\! \set{\!\l_{j,u}(\phi),\lD_{j,u}(i)\!} \leq  \min\! \set{\!\l_{j,u}(\phi),\lD_{j,u}(i)\!} + \frac{\tau_e}{2} 
\leq \min\! \set{\!\l_{j,v}(\phi), \lD_{j,v}(i)\!} - \frac{\tau_e}{2} \leq (K-1) \istep.
\end{equation}
For the first inequality note that we use $\sqrt{\alpha} < \istep C^{-\hat H} < \tau_e / 2 \cdot C^{-(K-1)}$.
By introducing the hypothetical flow on arc $e$, we can apply \Cref{lem:bound_on_exit_times} with $\delta \coloneqq
C^{K-1} \sqrt{\alpha}$. The preconditions of the lemma are satisfied since we can apply the induction hypothesis to
every $\varphi \in \beta \N$ (and $i' \coloneqq \varphi / \beta$) with $\min\set{\!\l_{j'\!,u}(\varphi), \lD_{j'\!,u}(i')\!} \leq \max\set{\!\l_{j,u}(\phi),
\lD_{j,u}(i)\!}$ thanks to \eqref{eq:induction_step_max}. Note that $k_{j'\!,u}(\phi')=\lD_{j'\!,u}(i')$ for all $i' \in
N_{j'}$ and $\phi'=i'\beta \in M_{j'}$. \Cref{lem:bound_on_exit_times} gives us
\[\abs{\l_{j,v}(\phi) -k_{j,v}(\phi)} \leq  11 \frac{\kappa_e}{\nu_e}  C^{K-1} \sqrt{\alpha}.\]
We have chosen $\alpha$ small enough such that $C^{K-1} \sqrt{\alpha} \leq \istep$. Hence, again by the induction
hypothesis, we obtain
\[\lD_{j,u}(i)\leq \min \set{\lD_{j,u}(i), \l_{j,u}(\phi)} + C^{K-1} \sqrt{\alpha}< (K-1) \istep + \istep\leq K \istep.\]
Applying \Cref{lem:arc_rounding_error} with $\theta \coloneqq K \istep$ yields
\[ \abs{k_{j,v}(\phi) -\lD_{j,v}(i)} \leq  C_1 \sqrt{\alpha} + K \istep C_2 \sqrt{\alpha}.\]
Finally, by choosing $C$ large enough, we obtain
\begin{align*}
\abs{\l_{j,v}(\phi) -\l^D_{j,v}(i)} &\leq \abs{\l_{j,v}(\phi) -k_{j,v}(\phi)} +\abs{k_{j,v}(\phi) -\lD_{j,v}(i)} \\
&\leq 11 \frac{\kappa_e}{\nu_e}  C^{K-1} \sqrt{\alpha} + C_1 \sqrt{\alpha} + K \istep C_2 \sqrt{\alpha}\\
&\leq (C-\sqrt{C}) C^{K-1} \sqrt{\alpha}  +K C \sqrt{\alpha}\\
&\leq \left(C^{K}  -  C^{K-0.5} + KC\right)\sqrt{\alpha}\\
& \leq C^K \sqrt{\alpha}.
\end{align*}
The inequalities hold by choosing $C$ large enough such that the following conditions are satisfied
\[C -\sqrt{C} \geq 11 \frac{\kappa_e}{\nu_e}, \qquad C \geq C_1 + C_2 \istep \quad \text{ and }\quad C^{K-0.5} 
\geq  KC\;\;\text{ for } K \in \set{2, 3, \dots, \hat H}.\]
Note that the last condition holds for all $C \geq 4$.
\end{proof}
\medskip

Since each commodity has a bounded flow mass and the supply rate functions have a bounded support, there exists a point
in time when the last particle enters the network. As the particles travel along predefined paths there exists a finite
time horizon, i.e., a point in time at which all particles have arrived at their destinations and the network is empty
from this point in time onward. By \Cref{thm:convergence} this implies that there is also a finite time horizon for all
packet routing instances with $\alpha \leq 1$ simultaneously. With this observation, \Cref{thm:convergence} implies
convergence of the arrival times, and even stronger, it is possible to show that the flows converge by proving uniform
convergence of the cumulative flow rates.

\begin{theorem} \label{thm:convergence_of_arrival_times_and_flows}
Let $(V, E, (\nu_e)_{e \in E}, (\tau_e)_{e \in E})$ be a network and $J$ be a finite set of commodities, each $j \in J$
equipped with a simple $o_j$-$d_j$-path $P_j$ and an integrable and bounded supply rate function $u_j$ with bounded
support. Consider the packet routings induced by a sequence of discretization parameters~$(\alpha, \beta)$ such that
$\alpha$ and $\frac{\beta}{\alpha}$ go to zero. Then the following two statements hold:
\begin{enumerate} \renewcommand\labelenumi{(\roman{enumi})}
\renewcommand\theenumi\labelenumi
\item For every $v \in V$, $j \in J_v$ and $\phi\in M_j$, the arrival time $\lD_{j,v}(\lceil\frac{\phi}{\beta}\rceil)$
of the packet model converges to the arrival time $\l_{j,v}(\phi)$ of the flow model. \label{it:convergence_of_arrival_time} 
\item For all $e = uv \in E$ and all $j \in J_e$ the cumulative flow $G_{j,e}^+$ of the packet model converges uniformly
to the cumulative flow $F_{j,e}^+$ of the flow model. Analogously, $G_{j,e}^-$ converges uniformly to $F_{j,e}^-$.
\label{it:convergence_of_cumulative_flows}
\end{enumerate}
\end{theorem}

\begin{proof}
For \ref{it:convergence_of_arrival_time} first observe that by the continuity of $\l_{j,v}$ we have that
$\l_{j,v}(\lceil \phi \rceil_\beta) \to \l_{j,v}(\phi)$ for $\beta \to 0$. The statement then follows immediately by
\Cref{thm:convergence} with $K$ large enough such that $\l_{j,v}(\phi) \leq K \istep$.

For \ref{it:convergence_of_cumulative_flows} let $\epsilon > 0$. By \Cref{thm:convergence} it is possible to choose
$(\alpha, \beta)$ small enough such that
\[\abs{\lD_{j, u}(i) - \l_{j, u}(i \beta)} \leq \frac{\epsilon-\beta}{\kappa_e}\]
for every packet $i \in N_j$. This is possible by again choosing $K$ large enough such that at time $K \istep$ all
particles have left the network.
Given a point in time $\theta$, we consider the particle $\phi \coloneqq F_{j,e}^+(\theta) \in M_j$. We denote the two
packets that are as close as possible to $\phi$ by  $i^- \coloneqq \lfloor\frac{\phi}{\beta}\rfloor \in N_j$ and $i^+
\coloneqq \lceil\frac{\phi}{\beta}\rceil \in N_j$ (see below for the case that one of them does not exist).
Note that the arrival time of packet $i^-$ cannot be much later than $\theta$, and analogously, the arrival time of
$i^+$ cannot be much earlier than $\theta$. More precisely,
\[\lD_{j,u}(i^-) - \frac{\epsilon-\beta}{\kappa_e} \leq \l_{j,u}(i^- \beta) \leq \theta \leq \l_{j,u}(i^+ \beta) 
\leq \lD_{j,u}(i^+) + \frac{\epsilon-\beta}{\kappa_e}.\]
Since the slope of $G_{j,e}^+$ is bounded by $\kappa_e$ and $F_{j,e}^+(\theta) \leq  i^- \beta + \beta$, we obtain
\[G_{j,e}^+(\theta) \geq G_{j,e}^+(\lD_{j,u}(i^-)) - \frac{\epsilon-\beta}{\kappa_e} \cdot \kappa_e  
= i^- \beta -(\epsilon-\beta) \geq F_{j,e}^+(\theta) - \epsilon. \]
Analogously, we get an upper bound on $G_{j,e}^+$ by using $F_{j,e}^+(\theta) \geq  i^+ \beta - \beta$
\[G_{j,e}^+(\theta) \leq G_{j,e}^+(\lD_{j,u}(i^+)) + \frac{\epsilon-\beta}{\kappa_e} \cdot \kappa_e
=i^+\beta + \epsilon-\beta \leq F_{j,e}^+(\theta) + \epsilon.\]
For the case that $i^- = \lfloor\frac{\phi}{\beta}\rfloor = 0 \not \in N_j$, we use the trivial lower bound of
$G_{j,e}^+(\theta) \geq 0$ and for the case that $i^+ = \lceil\frac{\phi}{\beta}\rceil = \abs{N_j} + 1 \not \in N_j$, we
use the upper bound $G_{j,e}^+(\theta) \leq \abs{N_j}\cdot\beta \leq \phi + \beta \leq  F_{j,e}^+(\theta) + \epsilon$.
With the exact same arguments, it holds that $\abs{G_{j,e}^-(\theta) - F_{j,e}^-(\theta)} \leq \epsilon.$
\end{proof}
\medskip
 
\subsection{Example.}

\begin{figure}[!b]
\centering
\includegraphics[width=\textwidth]{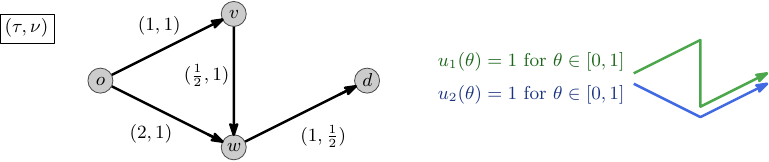}
\caption{Network, where the arc properties are given in the form $(\tau, \nu)$. At $w$, arc $ow$ is preferred over arc $vw$ in a tie break. We consider two commodities, one sending flow along the zigzag path $(o,v,w,d)$, one along the bottom path $(o,w,d)$. Both commodities send flow with a rate of one in $[0,1]$.}
\label{fig:network_convergence_example}
\vspace{1em}
\includegraphics[width=\textwidth]{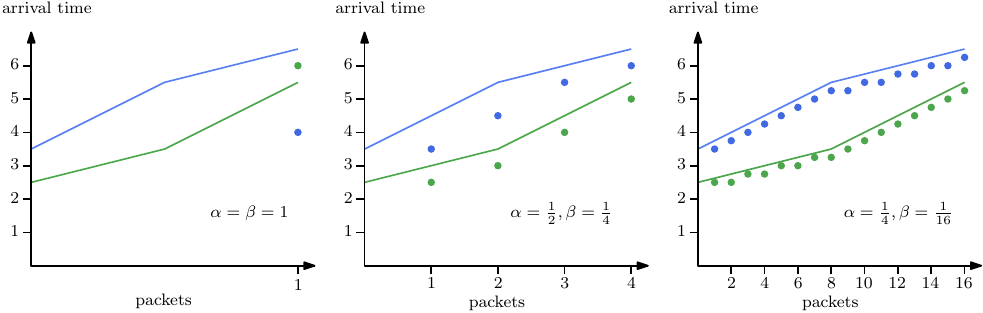}
\caption{The arrival times at the destination $d$ of the particles/packets of the different commodities (green for the first commodity along the zigzag path, blue for the second on the bottom path). The solid lines represent the arrival time functions of the continuous model. On the left, the packet size and the time step size are equal to one. In the middle the discretization is given by $\alpha=\frac{1}{2}, \beta=\frac{1}{4}$ and on the right by $\alpha=\frac{1}{4}, \beta=\frac{1}{16}$. The horizontal axis shows the indices of the packets. Overall the packets always sum up to a flow volume of $1$. The vertical axis corresponds to the arrival times.}
\label{fig:convergence_example}
\end{figure}

To complement this section, we want to visualize the convergence result in a simple example. For this we consider the network depicted in \Cref{fig:network_convergence_example}. 
The first commodity (green) sends one flow unit within $[0,1]$ with a rate of one along the zigzag path. The second commodity (blue) sends flow from time $0$ to $1$ with a rate of one along the bottom path.
We compare the arrival time of particles and packets for different discretizations in \Cref{fig:convergence_example}. It becomes apparent that for smaller discretization the arrival times approach the arrival time of the continuous flow embedding. 

In detail, the continuous flow of the green commodity reaches node $w$ half a time unit earlier than the flow of the blue commodity. Thus, the first half of the green flow (particles $[0,\frac{1}{2}]$) can use the full capacity of arc $wd$ and the second half splits capacity evenly with the blue commodity. Afterwards, all green flow particles have left $wd$ and the blue commodity can use the full capacity.

In the first packet routing instance with discretization $(1,1)$, we only consider the network at discrete time steps. This means at time $2$ both the single blue and the single green packet are located at $w$ and enter arc $wd$. Since the arc of the blue packet is preferred in a tie break, the blue packet enters arc $wd$ first, followed by the green packet (note that both packets enter the arc at the same time step, still the ordering of the packets is essential). Now, to leave the arc, and to arrive at node $d$, a capacity of one is needed. For that reason capacity has to be accumulated at $wd$ for one time step before the blue packet can leave the arc at time $4$. Again, capacity needs to be accumulated for one time step until the green packet leaves the arc at time $6$.

Already in the packet routing instance with discretization $(\frac{1}{2},\frac{1}{4})$, the packet arrival times become much more similar to the arrival times of the continuous model. First of all, since time steps have a length of $\frac{1}{2}$, the difference in the travel times from $o$ to $w$ along the two paths becomes noticeable and the first two green packets enter arc $wd$ before the first blue packet arrives at $w$. Second, a capacity of $\frac{1}{2}$ allows exactly one packet of size $\frac{1}{4}$ to leave every time step of length $\frac{1}{2}$, which makes the outflow of arc $wd$ smoother. 

In the packet routing instance with discretization $(\frac{1}{4}, \frac{1}{16})$, we have a very similar situation as in the continuous flow. The first half of the green packets can use arc $wd$ first. Afterwards capacity is split evenly until the last green packet has arrived at $d$ and the blue packets can use the full capacity for themselves.

\section{Existence of approximate pure Nash equilibria.} \label{sec:game_theory}
As an application of the convergence result, we show the existence of approximate pure Nash equilibria in our competitive
packet routing model by using the existence of exact dynamic equilibria for flows over time. This is particularly
interesting for the multi-commodity setting, where exact pure Nash equilibria do not exist in general as it was shown in
\Cref{thm:no_equilibria}.

\begin{theorem}
Consider a network $(V, E, (\nu_e)_{e \in E}, (\tau_e)_{e \in E})$, a constant $T > 0$, and a set of commodities~$J$,
each $j \in J$ equipped with an origin-destination-pair $(o_j, d_j)$ and an integrable and bounded supply rate function
$u_j \colon [0,T] \to \R_{\geq 0}$. For every $\epsilon>0$, there are discretization parameters $(\alpha,\beta)$, such
that the corresponding competitive packet routing game possesses an $\epsilon$-equilibrium.
\end{theorem}
\begin{proof}
Fix $\epsilon > 0$. We consider a multi-commodity Nash flow over time $f$, which exists due to Cominetti et
al.~\cite[Theorem 8]{cominetti2015existence}. It can be decomposed into a path-based flow over time $(f_P)_{P \in
\mathcal{P}}$, where 
\[\mathcal{P} \coloneqq \set{P \text{ is an $o_j$-$d_j$-path of some commodity $j$}}\] (cf.\
\cite[Section 2.7]{cominetti2015existence}). Additionally, Cominetti et al.\ have shown in~\cite[Lemma
7]{cominetti2015existence} that the path arrival times $(\l_{P,d_P})_{P \in \mathcal{P}}$ depend continuously on the
path inflow rates $(f_P)_{P \in \mathcal{P}}$. Hence, we can choose $\delta > 0$ small enough such that for all path
inflow rates $(\tilde{f}_P)_{P \in \mathcal{P}}$, which differ by at most $\delta$ from $(f_P)_{P \in \mathcal{P}}$
(i.e., $\sum_{P \in \P} \lVert f_P -\tilde{f}_P\rVert_{L^2} \leq \delta$) the corresponding arrival times $(\tilde
\l_{P, d_P})_{P \in \mathcal{P}}$ satisfy
\begin{equation}
\label{eq:continuity_of_arrival_times}
\abs{\l_{P,d_{P}}(\phi) - \tilde{\l}_{P,d_{P}}(\phi)} \leq \frac{\epsilon}{4}
\quad \text{ for all particles $\phi$ and all paths $P \in \mathcal{P}$.}
\end{equation}

By considering each path $P$ in $\mathcal{P}$ as a separate commodity, we obtain path-based supply rates $(u_P)_{P \in
\mathcal{P}} \coloneqq (f_P)_{P \in \mathcal{P}}$. Combining \Cref{thm:convergence} with a suitable time horizon $H$
provides a pair of discretization parameters $(\alpha,\beta)$ such that for every packet $i \in N_P$ in the
corresponding packet routing it holds that
\[\abs{\l_{P,d_P}(i \beta) - \lD_{P,d_P}(i)} \leq \frac{\epsilon}{4} \qquad \text{ for all } P \in \mathcal{P}.\]
Additionally, we require $\beta$ to be small enough such that \eqref{eq:continuity_of_arrival_times} holds for all
$\tilde{f}$ that can be created by shifting a flow volume of at most $\beta$ within $f$.

In the following, we show that the obtained packet routing constitutes an $\epsilon$-equilibrium. Assume for
contradiction that there is a packet $i$ that would arrive at least $\epsilon$ earlier by switching from its path $P$ to
some other $o_P$-$d_P$-path $P'$. By considering the modified supply rates $\tilde{u}$ that result from shifting the
$\beta$ flow volume corresponding to packet $i$ from $P$ to $P'$, we obtain a modified flow over time $\tilde{f}$ and a
modified packet routing with arrival times $\tilde{\l}$ and $\tilde{\l}^D$. Again with \Cref{thm:convergence}, we obtain
that
\[\abs{\tilde{\l}_{P',d_{P}}(i \beta) - \tilde{\l}^D_{P',d_{P}}(i)} \leq \frac{\epsilon}{4}.\]
Note that the constant $C$ in \Cref{thm:convergence} only depends on the instance. By choosing $C$ large enough, it is
possible to shift the flow volume without loosing the bound on the arrival times.

Combining all this, we obtain
\begin{align*}\l_{P,d_P}(i \beta) - \l_{P', d_{P'}}(i \beta) 
&\hspace{-1pt}\stackrel{\eqref{eq:continuity_of_arrival_times}}{\geq} \l_{P,d_P}(i \beta) - \tilde{\l}_{P',d_{P'}}(i \beta)
- \frac{\epsilon}{4}\\
&\geq \lD_{P,d_P}(i) - \tilde{\l}^D_{P',d_{P'}}(i) - \frac{3 \epsilon}{4}\\
&\geq \epsilon - \frac{3 \epsilon}{4} > 0.
\end{align*}
But this is a contradiction to the fact that $f$ is a Nash flow over time since all particles of the flow volume
corresponding to packet $i$ use path $P$ even though path $P'$ is strictly faster.
\end{proof}
\medskip

\section{Open problems.}
This work is the first that proves a connection between continuous dynamic models and discrete packet routing models.
There are several possible directions to proceed with this research.

Our bound on the discretization error increases exponentially in time.
Experiments~\cite{ZiemkeEtAl2020FlowsOverTimeAsLimitOfMATSim} indicate that there might be a time-independent bound
though. Can a better upper and maybe also a lower bound on the worst-case error be shown mathematically? Furthermore,
what is the best convergence rate (in $\alpha$) for a fixed particle? We prove a convergence rate of $\sqrt{\alpha}$,
which is probably not the best possible. However, even rounding the transit time of a single arc causes a deviation of up
to $\alpha/2$ showing that a convergence rate cannot be faster than linear in $\alpha$.

Our convergence result in \Cref{thm:convergence_of_arrival_times_and_flows} focuses on the embedding for fixed path choices of the packets/particles.
An interesting but probably challenging follow-up question is the convergence of equilibria in the packet routing model to Nash flows over time (the equilibria in the continuous flow model).
More precisely, given a sequence of exact packet routing equilibria, whose discretization
parameters converge to zero, does the limit exist? If yes, does it constitute a Nash flow over time? 
As equilibria do not exist in general for competitive packet routing games with multiple origin-destination pairs, it seems reasonable to focus on the single origin-destination case for these kinds of questions.
As a first step, a deeper understanding of Nash flows over time is necessary. One essential aspect is the robustness. If slight distortions of the flow over time or small perturbations of the network led to drastically different dynamic equilibria, there would be no hope for such a convergence result as packets (no matter how small they are) would always cause some inaccuracy. 
Fortunately, this essential condition has been studied by Olver et al.~\cite{OlverEtAl2022ContinuityUA} recently. The authors show continuity of Nash flows over time with respect to several parameters, such as transit times, capacities, and initial starting conditions. 
Still, given continuity, convergence does not follow easily.
It would be necessary to overcome the challenge that packets cannot split into smaller pieces and traverse multiple paths at once, which is one key aspect of Nash flows over time. 
Also, the robustness results in \cite{OlverEtAl2022ContinuityUA} are not sufficient as they focus on distortion at a single point in time. However, a packet flow deviates from a Nash flow over time at almost all points in time.
Further investigation in this regard would certainly be interesting.

In the transport simulation MATSim, traffic assignments are computed by an iterative best response procedure. Naturally,
this raises the question whether best response dynamics converge to approximate equilibria in the competitive packet
routing model. If this could be proven mathematically, it would further extend the theoretical foundation of MATSim.

\section*{Acknowledgments.}
We thank Tobias Harks, Kai Nagel, and Martin Skutella for fruitful discussions on the topic. We also thank the anonymous
reviewers for their very helpful comments and questions.

This work was funded by the Deutsche Forschungsgemeinschaft (DFG, German Research Foundation)
under Germany’s Excellence Strategy -- The Berlin Mathematics Research Center MATH+ (EXC-2046/1, project
ID: 390685689) and the Research Training Group 2236 UnRAVeL.

\bibliographystyle{abbrv}

\bibliography{literature} 

\end{document}